\documentclass[AEJ]{AEA}
\usepackage[spanish,english]{babel}
\usepackage{relsize}\usepackage{xcolor}
\usepackage{amsfonts,multirow,amsmath,adjustbox,natbib}
\usepackage{accents}\usepackage{booktabs}\usepackage{lscape}
\usepackage{animate}
\usepackage[utf8]{inputenc}
\usepackage{subcaption}
\usepackage{caption} 
\usepackage{longtable}
\usepackage{bm}
\usepackage{bbm}
\usepackage[textsize=tiny]{todonotes}
\makeatletter
\usepackage{pseudocode}\usepackage{algorithm}\usepackage{float}\usepackage{algpseudocode} 
\AtBeginDocument{
	\g@addto@macro{\appendix}{\renewcommand{\p@subsection}{}}
}
\makeatother
\usepackage{color,soul}
\DeclareMathOperator{\E}{\mathbb{E}}
\usepackage{soul,color}
\usepackage[para,online,flushleft]{threeparttable}\usepackage{multirow}\usepackage{multicol}
\usepackage{graphicx, comment}
\definecolor{ahjcolor}{rgb}{0.0, 0.13, 0.40}			
\usepackage[colorlinks	=true,										
linkcolor	=ahjcolor,									%
urlcolor	=ahjcolor,									%
citecolor	=ahjcolor,
bookmarksnumbered=true]{hyperref}
\draftSpacing{1.5}
\begin{document}
	
	\title{Inferring hidden potentials in analytical regions: uncovering crime suspect communities in Medell\'in}
	
	\author{Alejandro Puerta\thanks{Department of Economics, School of Economics and Finance, Universidad EAFIT, Medell\'{i}n, Colombia, \href{E-mail: apuerta5@eafit.edu.co}{E-mail: apuerta5@eafit.edu.co}} and Andr\'es Ramírez--Hassan\thanks{Department of Economics, School of Economics and Finance, Universidad EAFIT, Medellín, Colombia. Grupo de estudios en economía y empresa. \href{aramir21@eafit.edu.co}{E-mail: aramir21@eafit.edu.co}}}
	\date{\today}\pubMonth{}\pubYear{}\pubVolume{}\pubIssue{}
	\begin{abstract} 
		This paper proposes a Bayesian approach to perform inference regarding the size of hidden populations at analytical region using reported statistics. To do so, we propose a specification taking into account one-sided error components and spatial effects within a panel data structure. Our simulation exercises suggest good finite sample performance. We analyze rates of crime suspects living per neighborhood in Medellín (Colombia) associated with four crime activities. Our proposal seems to identify hot spots or ``crime communities'', potential neighborhoods where under-reporting is more severe, and also drivers of crime schools. Statistical evidence suggests a high level of interaction between homicides and drug dealing in one hand, and motorcycle and car thefts on the other hand. 
	\end{abstract}
	\maketitle
	\section{Introduction}\label{sec:1}
	We introduce in this paper an inferential framework within a Bayesian paradigm to incorporate that reported rates of individuals with unknown features in analytical regions might be lower bounds. For instance, positive rate tests of individuals having particular diseases, drug consumers, cheating behavior or police report rates of criminal activity in analytical regions are under reports of the total population size. All these are examples of what we called ``hidden populations''. Our proposal controls for spatial effects and unobserved heterogeneity on panel data settings.\par
	
	We depart from the stochastic frontier analysis (composed error models) \citep{Aigner1977,Meeusen1977}, and consider inefficiency, which is a one-sided error term from a statistical point of view, as the conditional percentage of the total population actually reported per analytical region. Accordingly, we incorporate structural and transitory lower bounds as one-sided error terms. In particular, we extend \cite{tsionas2014firm} proposal by including spatial effects; this component induces heteroscedasticity, and as a consequence, omitting one-sided error terms potentially causes biased estimators \citep{Wang2002}.\par 
	
	Crime and arrest rates at a regional level based on reported statistics are lower bounds \citep{Kirk2006}. Consistently, we apply our proposal to model the rate of crime suspects living per neighborhood in Medellín (Colombia). In particular, we include spatial effects as our cross-section units are neighborhoods, and permanent and transient one-sided errors to handle time-varying lower bound issues. In this way, we can perform inference about potential under reports of crime suspects living in a region; this would be valuable for policymakers. Particularly, it would suggest regional hot spots to implement structural interventions and efficient police operations to reduce criminality. This is especially relevant in a city like Medellín, which is a natural experiment field due to its history of violence associated with drug trafficking and urban wars. Nevertheless, it is worth mentioning that Medellín is also an exceptional example of social transformation based on public and private interventions.\par 
	
	It seems that mainstream crime literature neither has taken into account potential bias due to omitting lower bound issues nor has implemented strategies to infer the population of criminals. A remarkable exception for the latter is \cite{Heijden2014}, who based their strategy on count models, but requires data at the individual level, whereas ours requires data at an aggregated regional level. \cite{andresen2006spatial,kakamu2008spatial,kikuchi2010neighborhood,arnio2012demography,He2015} take into account spatial effects modeling crime data. However, these authors do not take the lower bound issue into account, which may have consequences on the statistical properties of estimators. Observe that unlike most of the crime literature, which focuses on crime rates, we model place of residence of crime suspects, that is, we consider that neighborhoods have socioeconomic conditions that might promote ``schools of criminality''. So, following the spirit of stochastic frontier analysis, we can consider these conditions as production factors of potential criminals.\par

	On the other hand, \cite{druska2003generalized,schmidt2009spatial,mastromarco2016modelling,tsionas2016spatial,glass2016spatial,gude2018heterogeneous} propose stochastic frontier models including spatial effects, but focusing on ``standard production functions'' such as GDP at state level. Additionally, the tendency is to incorporate both effects in the same error: spatial dependence and inefficiency, whereas we separate these components. Also, our proposal differs from previous stochastic frontier proposals with spatial effects due to including a random conditional autoregressive spatial (CAR) effect. \cite{schmidt2009spatial} includes CAR effects in the inefficiency component, whereas most of the stochastic frontier literature includes spatial autoregressive (SAR) components. The former proposal does not induce explicitly heteroscedasticity when inefficiency is omitted due to being present precisely in this omitted part. On the other hand, the SAR process is not Markovian, so it generates global spatial patterns, and it seems that criminal activity in Medellín is controlled by gangs with influence in specific areas \citep{Collazos2020}. Therefore, we use the CAR specification, which is a Markovian process in space \citep{ramirez2017welfare}, to control for local spatial effects.\par
	
	We aim to contribute to crime and stochastic frontier literature, but mainly, considering one minus the (exponential) one-sided errors as the percentage of covered (uncaptured) criminals helps to build a link between these two well-developed areas of knowledge.\par
	
	It seems from our simulation exercises that the sampling performance of our proposal is sound, allowing us to capture both the one-sided error terms and the spatial dependency. Additionally, it allows us to obtain good estimates for the location parameters in the presence of a five--way error component model. We also find that predictive inference regarding hidden populations has good predictive interval coverage.\par
	
	Our empirical analysis suggests that homicide and drug dealing are strongly linked as both seem to generate local urban displacement, share some determinants and same hot spots (crime communities), which are mainly located in the most west analytical region in Medell\'in. On the other hand, there are common links between car and motorcycle thefts. For instance, both crime activities are associated with high local unemployment rates, and there is a crime community in the central-west part of the city. Although, there are some isolated crime communities specialized in each crime activity with their own determinants.\par
	
	The remainder of this paper proceeds as follows: Section \ref{Sec:Eco} outlines our econometric model, the conditional posterior distributions, and the results of simulation exercises. Section \ref{Sec:App} presents our application. In particular, construction of the analytical regions, unconditional spatial analysis based on hypothesis tests using standardized rates, and posterior inferential results. Section \ref{Sec:Conc} concludes.
	
	\section{Econometric approach}\label{Sec:Eco}
	
	\subsection{The model}
	
	The point of departure is the observed under-reported ratio of the number of target individuals per inhabitants (${Y}_{it}$) at analytical region $i=1,2,\dots,N$ and time period $t=1,2\dots,T$, where target individuals belong to the ``hidden population'' ($P_{it}$), which is also standardized,  
	\begin{align}\label{eq:sim0}
	{Y}_{it} & = P_{it} \times \text{R}_{it},
	\end{align}
	
	where $\text{R}_{it}$ is the report rate, that is, the percentage of individuals belonging to the target population that have been observed.
	
	We can think about $P_{it}$ as depending on environmental variables that promote or discourage the number of individuals belonging to the target population as well as spatial effects reflecting spatial clusters, and also unobserved regional heterogeneity and idiosyncratic stochastic errors. So, we propose $P_{it}=f(\bm X_{it},\bm \beta)\times \exp\{\alpha_i+v_i+\epsilon_{it}\}=\prod_{k=1}^K X_{kit}^{\beta_k}\times \exp\{\alpha_i+v_i+\epsilon_{it}\}$ where $X_{kit}$ are $k$ potential drivers which may include spatial lags, that is, given a set of controls ($\bm{z}_{it}$), their spatial lags are $\sum_{j=1}^N w_{ij}\bm{z}_{ijt}$, where $w_{ij}$ is the $ij$-th element of the contiguity matrix $\mathbf{W}_N$, $\bm{x}_{it}=\left[\bm{z}_{it}' \:  \left(\sum_{j=1}^N w_{ij}\bm{z}_{ijt}\right)' \right]'$. The location parameters are given by $\beta_k$. In addition, $\alpha_i$ is the unobserved stochastic heterogeneity, $v_{i}$ is the spatial random effect, and $\epsilon_{it}$ is the idiosyncratic stochastic error.
	
	On the other hand, we specify the report rate as $\text{R}_{it}=\exp\{-\eta_{i}^{+}-u_{it}^{+}\}$ where $\eta_i^+$ and $u_{it}^+$ are one-sided positive stochastic errors to account for unobserved persistent and transient lower bound issues, that is, if $\eta_{i}^{+}=u_{it}^{+}=0$, then $R_{it}=1$, and $Y_{it}=P_{it}$, otherwise we observe just a lower bound of $P_{it}$. 
	
	Observe that this setting follows the statistical framework of stochastic frontier analysis with permanent and transient one-sided components, unobserved heterogeneity and spatial effects. Therefore, we extend \cite{tsionas2014firm} proposal including spatial effects,      
	
	\begin{align} \label{eq:1}
	y_{it} = \bm{x}_{it}'\bm{\beta}+\alpha_i+v_{i}-\eta_{i}^{+}-u_{it}^{+} +\epsilon_{it},
	\end{align}
	such that the reduce form equation \ref{eq:1} is in $\log$-$\log$ form.\par

	Following \cite{tsionas2014firm} we assume that the i.i.d random components have the following distributions:
	\begin{align}\label{eq:2}
	\alpha_{i}\sim \mathcal{N}(0,\sigma^2_\alpha), \quad \eta_{i}^+\sim \mathcal{N}^+(0,\sigma^2_\eta), \quad u_{it}^+\sim \mathcal{N}^+(0,\sigma^2_u) , \quad  \epsilon_{it}\sim \mathcal{N}(0,\sigma^2_\epsilon). 
	\end{align}
	We assume that each $v_{i}$ has an improper (intrinsic) conditionally autoregressive structure \citep{besag1991}:
	\begin{align}\label{eq:3}
	v_i|\bm{v}_{i\sim j}\sim \mathcal{N}\bigg(\sum_{i\sim j}\frac{w_{ij}v_j}{\sum_{i\sim j}w_{ij}},\frac{\sigma^2_v}{\sum_{i\sim j}w_{ij}}\bigg),
	\end{align}
	where $\bm{v}_{i\sim j}$ is a vector of stochastic spatial errors for the neighbors $j$ of $i$ ($i\sim j$).\par
	
	The joint distribution of the improper CAR is $\bm{v}\sim \mathcal{N}(\bar{\bm{v}},\sigma^2_v(\bm{D}_w-\bm{W}_N)^{-1})$, where $\bm{D}_w=\text{diag}(\sum_{i\sim j}w_{ij})$ \citep{banerjee2014hierarchical}. The $ij$-th element of $\bm{W}_N$ is equal 1 if region $i$ and $j$ are neighbors, and 0 otherwise. By definition the elements of the main diagonal are set equal to zero.
	
	\subsection{Likelihood and priors}
	Set $\tau_{it}=\alpha_i+\epsilon_{it}$ such that taken assumptions in (\ref{eq:2}) into account, $\bm{\tau}_i\sim \mathcal{N}(\bm{0},\bm{\Sigma})$, $\bm{\Sigma}=\sigma_{\epsilon}^2\bm{I}_T+\sigma_{\alpha}^2\bm{i}_T\bm{i}_T'$, where $\bm{i}_T$ is a $T$-dimensional vector of 1's and $\bm{I}_T$ is a $T$-dimensional identity matrix. Given our model specification (equations (\ref{eq:1}), (\ref{eq:2}) and (\ref{eq:3})), the joint conditional distribution function, given spatial random effects, is the product over individuals of a $T$-variate closed skew normal distributions \citep{Dominguez2003}. Working directly with this distribution is demanding given that it is not readily available in closed form. So, we follow \cite{Sanchez2020} who in similar settings use data augmenting \citep{Tanner1987}. In particular, set $\bm{\theta}=(\bm{\bm{\beta}}',\sigma^2_\alpha,\sigma^2_\epsilon,\sigma^2_v,\sigma^2_u,\sigma^2_\eta)'$, and the augmented vector $\bm{\Theta}=(\bm{\theta}',u_{it}^+,\eta_{i}^+,v_i)$, then taking into account equations (\ref{eq:1}), (\ref{eq:2}) and (\ref{eq:3}), the ``augmented'' likelihood is 
	\footnotesize
	\begin{align*}
	f(\bm{y}|\bm{X},\bm{\Theta})&=\prod_{i=1}^{N}(2\pi)^{-\frac{T}{2}}|\bm\Sigma|^{-\frac{1}{2}}\text{exp}\biggl\{-\frac{1}{2}\big(\bm{y}_i-\bm{X}_i\bm{\bm{\beta}}-\bm{u}_{i}^+-v_i\bm{i}_T-\eta_i^+\bm{i}_T\big)'\bm\Sigma^{-1}\big(\bm{y}_i-\bm{X}_i\bm{\bm{\beta}}-\bm{u}_{i}^+-v_i\bm{i}_T-\eta_i^+\bm{i}_T\big)\biggr\} \nonumber\\
	&\times
	(2\pi)^{-\frac{T}{2}}\bigg(\frac{\sigma_v^2}{\sum_{i\sim j}w_{ij}}\bigg)^{-\frac{T}{2}}\text{exp}\biggl\{-\frac{\sum_{i\sim j}w_{ij}}{2\sigma_v^2}\bigg(v_i\bm{i}_T-\sum_{i\sim j}\frac{w_{ij}v_j}{\sum_{i\sim j}w_{ij}}\bm{i}_T\bigg)'\bigg(v_i\bm{i}_T-\sum_{i\sim j}\frac{w_{ij}v_j}{\sum_{i\sim j}w_{ij}}\bm{i}_T\bigg)\biggr\}\nonumber\\
	&I(\bm{u}_i^+>0)\bigg(\frac{2}{\pi}\bigg)^{-\frac{T}{2}}(\sigma^2_u)^{-\frac{T}{2}}\text{exp}\biggl\{-\frac{1}{2\sigma_u^2}\bm{u}_i^{+'}\bm{u}_i^{+}\biggr\}\nonumber\\
	&I(\eta_i^+>0)\bigg(\frac{2}{\pi}\bigg)^{-\frac{1}{2}}(\sigma^2_\eta)^{-\frac{1}{2}}\text{exp}\biggl\{-\frac{1}{2\sigma_\eta^2}\eta_i^{+2}\biggr\},
	\end{align*}
	\normalsize
	where we stack information by individual such that $\bm{X}_i$ is a $T\times dim\left\{\bm\beta\right\}$ dimensional matrix with information of individual $i$.\par
	
	We follow standard practice in Bayesian econometrics with conditional independent priors such that $\bm{\bm{\beta}}\sim \mathcal{N}(\bm{\beta}_0,\bm{B}_0)$, where $\bm{\beta}_0=\bm{0}$ and $\bm{B}_0=1000\bm{I}$ which implies vague prior information. For the scale parameters, 
	$\frac{\bar{Q}_k}{\sigma^2_k}\sim \chi^2(\bar{N}_k)$, $k=\epsilon,\alpha,v$, where $\bar{N}_k=1$ and $\bar{Q}_k=10^{-4}$ \citep{tsionas2014firm}. We follow \cite{makiela2017bayesian} for the priors of $\sigma_u^2$ and $\sigma_\eta^2$, that is, we use $\sigma_u^{2}\sim \mathcal{I}\mathcal{G}(v_{0u}/2,2v_{0u}\log^2(r_u^*)/2)$ and $\sigma_{\eta}^{2}\sim \mathcal{I}\mathcal{G}(v_{0\eta}/2,2v_{0\eta}\log^2(r_{\eta}^*)/2)$ where the prior medians of the transient and persistent one-sided errors are equal to $r^*_u=0.85$ and $r^*_{\eta}=0.70$, and $v_{0u}=v_{0\eta}=10$.\footnote{We perform robustness analysis regarding these hyperparameters. Available upon request.}
	Even though $\bm{v}$ has an improper distribution, Theorem 2 in \cite{sun1999posterior} guarantees that a proper posterior distribution exists if $\bm{D}_w-\bm{W}_N$ is nonnegative definite, the precision parameters have gamma prior distributions, and the intercepts have diffuse prior distributions (we fulfill all these requirements).
	
	\subsection{Conditional posterior distributions}\label{subsec:c}
	
	The conditional posterior distribution for the location parameters is
	\begin{align*}\bm{\beta}|\bm{\Theta}_{-\bm{\beta}},\bm{y},\bm{X}\sim N(\bar{\bm{\beta}},\bar{\bm B}),
	\end{align*}
	where $\bar{\bm B}=\bigg(\sum_i \bm{X}_i'\bm{\Sigma}^{-1}\bm{X}_i+\bm{B}_0^{-1}\bigg)^{-1}$, $\bar{\bm{\beta}}=\bar{\bm{B}}\bigg(\sum_i \bm{X}_i'\bm{\Sigma}^{-1}\tilde{\bm y}_i+\bm{B}_0^{-1}\bm{\beta}_0\bigg)$, and $\tilde{\bm y}_i=\bm{y}_i-\bm{u}_i^+ -v_i\bm{i}_T-\eta_i^+\bm{i}_T$. The notation $\bm\Theta_{-\psi}$ indicates all elements in $\bm\Theta$ except $\psi$.\par
	The conditional posterior distribution for $\bm{u}_i^+$ is
	\begin{align*}
	\pi(\bm{u_i}^+|\bm{\Theta}_{-\bm{u}_i^+},\bm{y},\bm{X})\propto I(\bm{u}_i^+>0)\times \text{exp}\biggl\{-\frac{1}{2}(\bm{u}_i^+-\bm{\mu})'\bm{\Omega}^{-1}(\bm{u}_i^+-\bm{\mu})\biggr\},
	\end{align*}
	where $\bm{\Omega}=\left(\bm{\Sigma}^{-1}+\bm{I}_T\frac{1}{\sigma_{u}^{2}}\right)^{-1}$ and $\bm{\mu}=\bm{\Omega}\bigg(\bm{\Sigma}^{-1}(\bm{y}_i-\bm{X}_i\bm{\beta}-v_i\bm{i}_T-\eta_i^+\bm{i}_T)\bigg)$. Given the multivariate condition $I(\bm{u}_i^+>0)$, which can be difficult to meet in high dimensional settings, we sample from $\pi(u_{it}^+|\bm{\Theta}_{-\bm{u}_{it}^+},\bm{y},\bm{X})$ using the result from a conditional multivariate normal distribution 
	\citep{eaton1983multivariate}. Let $\bm{u_i}^+=(u^+_{i1},\hdots,u^+_{iT})'$, \begin{align*}\bm{u_i}^+=\begin{pmatrix}
	u^+_{1t}  \\ \bm{u}^+_{2t}
	\end{pmatrix}\sim I(\bm{u}_i^+>0)\times\mathcal{N}\begin{bmatrix}
	\begin{pmatrix}
	\mu_{1}\\ \bm{\mu}_{2}
	\end{pmatrix},\begin{pmatrix}
	\omega_{11} & \bm{\Omega}_{12}  \\ \bm{\Omega}_{21}& \bm{\Omega}_{22} 
	\end{pmatrix}
	\end{bmatrix},
	\end{align*}
	then, the conditional distribution of $u^+_{1t}$ given $\bm{u}^+_{2t}$ is
	\begin{align*}
	u_{1t}^+|\bm{\Theta}_{-{u}_{1t}^+},\bm{y},\bm{X},\bm{u}^+_{2,t}\sim I(u_{1t}^+>0)\times \mathcal{N}(\bar{\mu},\bar{\omega}),
	\end{align*}
	where $\bar{\mu}=\mu_{1}+\bm{\Omega}_{12}\bm{\Omega}_{22}^{-1}(\bm{u}_{2t}^+-\bm\mu_{2})$ and $\bar{\omega}=\omega_{11}-\bm{\Omega}_{12}\bm{\Omega}_{22}^{-1}\bm{\Omega}_{21}$.\par
	
	The conditional posterior distribution for $\eta_i^+$ is:
	\begin{align*}
	\eta_i^+\sim \mathcal{N}^+(m_i,\psi^2)
	\end{align*}
	where $\psi^2=\sigma_\eta^2 \bigg(1+\sigma_\eta^2\bm{i}_T'\Sigma^{-1}\bm{i}_T\bigg)^{-1}$ and $m_i=\psi^2\bm{i}_T'\Sigma^{-1}(\bm{y}_i-\bm{X}_i\bm{\bm{\beta}}-\bm{u}_{i}^+-v_i\bm{i}_T)$. \par 
	The conditional posterior distribution for $v_i$ is
	\begin{align*}
	v_i|\bm{\Theta}_{-v_{i}},\bm{y},\bm{X}\sim N(\bar{v}_i,\bar{\sigma}^2_{vi}), 
	\end{align*}
	where $\bar{\sigma}^2_{vi}=\bigg(\bm{i}_{T}'\bm\Sigma^{-1}\bm{i}_{T}+\frac{\sum_{i\sim j}w_{ij}}{\sigma^2_v}\bigg)^{-1}$ and $\bar{v}_i=\bar{\sigma}^2_{vi}\bigg(\bm{i}_{T}'\bm\Sigma^{-1}(\bm y_{i}-\bm{X}_{i}\bm{\beta}-\bm{u}_{i}^+-\eta_i^+\bm{i}_T)+\sum_{i\sim j}\frac{w_{ij}v_j}{\sigma^2_v}\bigg)$.
	The conditional posterior distribution for $\sigma^2_v$ is
	\begin{align*}
	\left.\frac{\bar{Q}_v+\bm{v}'\big(\bm{D}_w-\bm{W}_N\big)\bm{v}}{\sigma^2_v}\right\vert\bm{\Theta}_{-\sigma_v},\bm{y},\bm{X} \sim \chi^2(N\times T+\bar{N}_v).
	\end{align*} 
	The conditional posterior distribution for $\sigma_{u}^2$ is
	\begin{align*}
	\sigma^2_u|\bm\Theta_{-\sigma_u},\bm{y},\bm{X} \sim \mathcal{I}\mathcal{G}\bigg(\frac{(N\times T)+v_{0u}}{2},\frac{\bm{u}^{+'}\bm{u}^{+}+2v_{0u}\log^2(r_u^*)}{2}\bigg).
	\end{align*}
	The conditional posterior distribution for $\sigma_{\eta}^2$ is
	\begin{align*}
	\sigma^2_\eta|\Theta_{-\sigma_\eta},\bm{y},\bm{X} \sim \mathcal{I}\mathcal{G}\bigg(\frac{N+v_{0\eta}}{2},\frac{\bm{\eta}^{+'}\bm{\eta}^+ + 2v_{0\eta}\log^2(r_{\eta}^*)}{2}\bigg).
	\end{align*}
	
	Up to this point we have standard conditional posterior distributions, so we can use the Gibbs sampling algorithm. However, the conditional posterior distributions for $\sigma_{\alpha}^2$ and $\sigma_{\epsilon}^2$ do not have standard form. We use the Metropolis-Hastings algorithm for these two parameters.  \par
	
	In particular, the conditional posterior distribution for $\sigma_{\alpha}^2$ is
	\small
	\begin{align*}
	\pi(\sigma_{\alpha}^2|\bm\Theta_{-\sigma_{\alpha}^2},\bm{y},\bm{X})\propto \prod_{i=1}^{N}|\big(\sigma^2_\epsilon\bm{I}_t+\sigma_\alpha^2\bm{i}_T\bm{i}_T'\big)|^{-\frac{1}{2}}\text{exp}\biggl\{-\frac{1}{2}\big(\bm{y}_i-\bm{X}_i\bm{\beta}-\bm{u}_i^+-v_i\bm{i}_T-\eta_i^+\bm{i}_T\big)'\\ 
	\big(\sigma^2_\epsilon\bm{I}_T+\sigma_\alpha^2\bm{i}_T\bm{i}_T'\big)^{-1} 
	\big(\bm{y}_i-\bm{X}_i\bm{\beta}-\bm{u}_i^+-v_i\bm{i}_T-\eta_i^+\bm{i}_T\big)\biggr\}
	\frac{1}{(\sigma_\alpha^2)^{\bar{N}_\alpha/2+1}}\text{exp}\biggl\{-\frac{\bar{Q}_\alpha}{2\sigma_\alpha^2}\biggr\}.	
	\end{align*}
	\normalsize
	The conditional posterior distribution for $\sigma_{\epsilon}^2$ is
	\small
	\begin{align*}
	\pi(\sigma_{\epsilon}^2|\bm\Theta_{-\sigma_{\epsilon}^2},\bm{y}, \bm{X})\propto \prod_{i=1}^{N}|\big(\sigma^2_\epsilon\bm{I}_t+\sigma_\alpha^2\bm{i}_T\bm{i}_T'\big)|^{-\frac{1}{2}}\text{exp}\biggl\{-\frac{1}{2}\big(\bm{y}_i-\bm{X}_i\bm{\beta}-\bm{u}_i^+-v_i\bm{i}_T-\eta_i^+\bm{i}_T\big)'\\ 
	\big(\sigma^2_\epsilon\bm{I}_T+\sigma_\alpha^2\bm{i}_T\bm{i}_T'\big)^{-1} 
	\big(\bm{y}_i-\bm{X}_i\bm{\beta}-\bm{u}_i^+-v_i\bm{i}_T-\eta_i^+\bm{i}_T\big)\biggr\}
	\frac{1}{(\sigma_\epsilon^2)^{\bar{N}_\epsilon/2+1}}\text{exp}\biggl\{-\frac{\bar{Q}_\epsilon}{2\sigma_\epsilon^2}\biggr\}.	
	\end{align*}
	\normalsize
	We use as proposal distributions scaled Chi-squared distributions with one degree of freedom. 
	
	\subsection{Simulations}
	
	We consider the following data generating process:
	\begin{align}\label{dgp}
	y_{it} = 0.5z_{it}+0.5\sum_{j=1}^N w_{ij}z_{ij,t}+\alpha_i+v_{i}-\eta_{i}^{+}-u_{it}^{+} +\epsilon_{it},
	\end{align}
	where the contiguity criterion is queen, and $z_{it}$ is drawn from a standard normal distribution. Following \cite{tsionas2014firm},  $\sigma_\alpha=0.1,\sigma_\eta=0.5,\sigma_u=0.2,$ and $\sigma_\epsilon=0.1$, and we set $\sigma_v=0.4$. We perform 20,000 iterations, a burn-in equal to 10,000, and a thinning parameter equal 5.\par 
	\subsubsection{Population parameters: point estimate results}
	Table \ref{tab:1} displays sampling properties regarding point estimates of our Bayesian proposal using different combinations of $N$ an $T$, where one of them closely matches our application. We can see that our Bayesian proposal has good sampling properties as the highest density intervals (HDI) are relatively narrow and contain the population scale and location parameters. Comparing the first scenario ($N=49$, $T=5$) with the last scenario ($N=196$, $T=10$), we observe that the HDIs get narrower as the sample size increases. This is particularly relevant for the scale parameters. In general, the expected value of the correlation of the posterior draws of one-sided errors and spatial effects, and the \textit{Population} values ($\hat{\mathbf{E}}(\rho)$) is higher than 0.5 except in one case. Also, the descriptive statistics (mean and median) of these unobserved stochastic components are similar.\par
	
	We produce another two sets of simulation results (see Appendix \ref{Appendix}, Tables \ref{tab:a1} and \ref{tab:a2}). The first shows the consequences of varying $\lambda=\frac{\sigma_{\eta}+\sigma_u}{\sigma_{\epsilon}}$. This parameter has attracted a lot of attention in the stochastic frontier community. \cite{olson1980monte} identifies two main issues when $\lambda \to \infty$: two-step estimators have very unstable empirical moments, and negative bias (constant term and $\hat{\sigma}^2_u$). On the other hand, the \textit{wrong skew} problem, $\lambda \to 0$, implies opposite direction bias. \cite{waldman1982stationary,horrace2019stationary} prove the existence of a stationary point in this case, then the probability of the \textit{wrong skew} problem converges to zero when the sample size converges to infinity. However, \cite{simar2009inferences} show that finite sample problems remain.\par
	Robustness checks for $\lambda$ are reported in Table \ref{tab:a1} in the Appendix. The sampling properties of our proposal seems to follow previous studies, that is, good performance as far as $\lambda$ is higher than one but bounded. Posterior estimates in our application suggests $\lambda$ is between 2 and 9, which seems a safe ground.\par
	
	We also perform robustness checks regarding the presence of heavy tails in the stochastic error ($\epsilon_{it}$). In particular, we assume normality to perform statistical inference, but simulating equation \ref{dgp} using a Student's t-distribution with four degrees of freedom. It seems from results in Table \ref{tab:a2} in the Appendix that our inferential procedure is robust to heavy tails presence.\par
	\subsubsection{Hidden population: coverage results}
	One of the main purposes of our proposal is being able to make inference about ``hidden populations''. Given $Y_{it}=P_{it}\exp\left\{-(\eta_i^+ + u_{it}^+)\right\}$, then $\left\{-(\eta_i^+ + u_{it}^+)\right\}$ is the percentage of reported cases of our target population, and as a consequence, 1-$\exp(-(\eta_i^+ + u_{it}^+))$ represents the percentage that is still covered. Therefore, we would expect that a sensible $1-\alpha$ predictive interval for the target population ($P_{it}$) is $\mathcal{P}_{\alpha}=\left\{Y_{it}\times\exp(\eta_i^+ + u_{it}^+):\pi(Y_{it}\times\exp(\eta_i^+ + u_{it}^+)|\bm\Theta,\bm Y, \bm X)\geq k(\alpha)\right\}$, $k(\alpha)$ is the largest constant such that $P(\mathcal{P}_{\alpha})\geq 1-\alpha$, that is, the $(1-\alpha)$ highest density (predictive) interval (HDI).
	\begin{landscape}
		\begin{table}[p] \caption{Sampling properties of Bayes estimators}\label{tab:1}
			\centering  \resizebox{1.48\textheight}{!}{%
				\begin{tabular}{cccccccccccccccccccccccccccccc}
					\hline    &
					\multicolumn{3}{c}{$\eta^+$} &
					\multicolumn{3}{c}{$\sigma_\eta$}&
					\multicolumn{3}{c}{$u^+$}&
					\multicolumn{3}{c}{$\sigma_u$} &
					\multicolumn{2}{c}{$v$}&
					\multicolumn{3}{c}{$\sigma_v$} &
					\multicolumn{3}{c}{$\sigma_\alpha$} &
					\multicolumn{3}{c}{$\sigma_\epsilon$} &
					\multicolumn{3}{c}{$\beta_1$} &
					\multicolumn{3}{c}{$\beta_2$} \\
					\cmidrule(r){2-4}  \cmidrule(r){5-7}  \cmidrule(r){8-10}  \cmidrule(r){11-13}  \cmidrule(r){14-15}  \cmidrule(r){16-18}  \cmidrule(r){19-21}  \cmidrule(r){22-24}  \cmidrule(r){25-27}  \cmidrule(r){28-30}
					&
					Mean &
					Median &
					$\hat{\E}(\rho)$    & Mean &
					\multicolumn{2}{c}{HDI}  &   Mean &
					Median &
					$\hat{\E}(\rho)$    & Mean &
					\multicolumn{2}{c}{HDI} & 
					Median &
					$\hat{\E}(\rho)$    & Mean &
					\multicolumn{2}{c}{HDI} & Mean &
					\multicolumn{2}{c}{HDI} & Mean &
					\multicolumn{2}{c}{HDI} & Mean &
					\multicolumn{2}{c}{HDI}& Mean &
					\multicolumn{2}{c}{HDI}  \\ 
					\hline
					Population & -0.367 & -0.314 &  & 0.500 &  &  & -0.152 & -0.132 &  & 0.200 &  &  & 0.005 &  & 0.400 &  &  & 0.100 &  &  & 0.100 &  &  & 0.500 &  &  & -0.500 &  &  \\ 
					Estimate & -0.362 & -0.262 & 0.744 & 0.478 & 0.387 & 0.576 & -0.164 & -0.135 & 0.527 & 0.209 & 0.156 & 0.249 & -0.028 & 0.558 & 0.364 & 0.185 & 0.513 & 0.057 & 0.003 & 0.131 & 0.092 & 0.059 & 0.131 & 0.493 & 0.471 & 0.517 & -0.496 & -0.505 & -0.487 \\ 
					N=49, T=5 &  &  &  &  &  &  &  &  &  &  &  &  &  &  &  &  &  &  &  &  &  &  &  &  &  &  &  &  &  \\ 
					Population & -0.338 & -0.266 &  & 0.500 &  &  & -0.172 & -0.145 &  & 0.200 &  &  & -0.031 &  & 0.400 &  &  & 0.100 &  &  & 0.100 &  &  & 0.500 &  &  & -0.500 &  &  \\ 
					estimate & -0.355 & -0.316 & 0.564 & 0.455 & 0.373 & 0.556 & -0.158 & -0.133 & 0.590 & 0.199 & 0.164 & 0.232 & 0.008 & 0.726 & 0.566 & 0.366 & 0.781 & 0.065 & 0.003 & 0.156 & 0.096 & 0.073 & 0.120 & 0.509 & 0.494 & 0.524 & -0.498 & -0.503 & -0.492 \\ 
					N=49, T=10 &  &  &  &  &  &  &  &  &  &  &  &  &  &  &  &  &  &  &  &  &  &  &  &  &  &  &  &  &  \\ 
					Population & -0.391 & -0.328 &  & 0.500 &  &  & -0.154 & -0.120 &  & 0.200 &  &  & 0.005 &  & 0.400 &  &  & 0.100 &  &  & 0.100 &  &  & 0.500 &  &  & -0.500 &  &  \\ 
					estimate & -0.374 & -0.289 & 0.707 & 0.487 & 0.407 & 0.569 & -0.183 & -0.153 & 0.588 & 0.231 & 0.172 & 0.275 & -0.005 & 0.573 & 0.447 & 0.280 & 0.608 & 0.073 & 0.003 & 0.150 & 0.086 & 0.045 & 0.134 & 0.493 & 0.477 & 0.507 & -0.497 & -0.503 & -0.490 \\ 
					N=100, T=5 &  &  &  &  &  &  &  &  &  &  &  &  &  &  &  &  &  &  &  &  &  &  &  &  &  &  &  &  &  \\ 
					Population & -0.396 & -0.291 &  & 0.500 &  &  & -0.163 & -0.137 &  & 0.200 &  &  & 0.005 &  & 0.400 &  &  & 0.100 &  &  & 0.100 &  &  & 0.500 &  &  & -0.500 &  &  \\ 
					estimate & -0.388 & -0.305 & 0.687 & 0.490 & 0.414 & 0.564 & -0.173 & -0.148 & 0.611 & 0.218 & 0.184 & 0.250 & 0.000 & 0.352 & 0.354 & 0.175 & 0.519 & 0.147 & 0.058 & 0.214 & 0.092 & 0.070 & 0.120 & 0.493 & 0.482 & 0.503 & -0.501 & -0.504 & -0.497 \\ 
					N=100, T=10 &  &  &  &  &  &  &  &  &  &  &  &  &  &  &  &  &  &  &  &  &  &  &  &  &  &  &  &  &  \\ 
					Population & -0.399 & -0.312 &  & 0.500 &  &  & -0.162 & -0.137 &  & 0.200 &  &  & 0.008 &  & 0.400 &  &  & 0.100 &  &  & 0.100 &  &  & 0.500 &  &  & -0.500 &  &  \\ 
					estimate & -0.416 & -0.358 & 0.735 & 0.521 & 0.461 & 0.581 & -0.154 & -0.134 & 0.542 & 0.194 & 0.153 & 0.224 & -0.012 & 0.565 & 0.474 & 0.350 & 0.600 & 0.081 & 0.006 & 0.144 & 0.099 & 0.076 & 0.125 & 0.504 & 0.498 & 0.511 & -0.500 & -0.503 & -0.498 \\ 
					N=196, T=10 &  &  &  &  &  &  &  &  &  &  &  &  &  &  &  &  &  &  &  &  &  &  &  &  &  &  &  &  &  \\ 
					\hline
			\end{tabular}} 
			\begin{tablenotes}
				\item \tiny \textit{Note: By \textit{Population} value of $\eta^+$, $u^+$ and $v$ we mean the \textit{average} and \textit{median} values as they were generated from the Monte Carlo experiment. We report posterior means, medians and highest density intervals from posterior draws using 20,000 iterations, 10,000 burn-in and thinning equal to 5. $\hat{\mathbf{E}}(\rho)$ is the expected value of the correlation of the posterior draws of one-sided errors and spatial effects, and the \textit{Population} one-sided errors and spatial effects coming from the Monte Carlo experiment, respectively.}
			\end{tablenotes}
		\end{table} 
	\end{landscape}
	\par We check the performance of this predictive interval calculating its coverage ($cov$). In particular, we propose a Beta-Binomial model for this coverage using as prior a non-informative Beta distribution, that is, $\pi(cov)\sim B(1,1)$, therefore its posterior distribution is $cov|\bm\Theta,\bm Y,\bm X\sim B(a,b)$ where $a=1+\sum_{i=1}^N\sum_{t=1}^T I_{it}$, $b=1+N\times T-\sum_{i=1}^N\sum_{t=1}^T I_{it}$,  $I_{it}=\mathbbm{1}\left[P_{it}\in \mathcal{P}_{\alpha}\right]$. Observe that $\mathbb{E}(cov|\bm\Theta,\bm Y,\bm X)\approx \frac{\sum_{i=1}^N\sum_{t=1}^T I_{it}}{N\times T}$ due to using a non-informative prior. Algorithm \ref{alg:power test} shows details.  
	
	It seems from Table \ref{tab:MC} that our proposal has a good coverage as all means are close to $(1-\alpha)$ credibility levels with narrow 95\% HDIs, where most of them embrace the credibility levels.\par 
	To have an idea of how close is our estimate of $P_{it}$ to the real value, we calculate the mean absolute percentage error (MAPE)  for each observation as
	\begin{align*}
	\text{MAPE}_{it}=\frac{1}{S}\sum_{s=1}^S\left|\frac{P_{it}-\widehat{\mathbb{E}}\left[P_{it}\right]}{P_{it}} \right|
	\end{align*}
	Table \ref{tab:cov} reports summary statistics corresponding the MAPE's for each of the sample sizes of Table  \ref{tab:MC}. Results suggest that our point estimate for $P_{it}$ tends to be very close to the real values. On average, our estimate diverges from $P_{it}$ in 23 percentage points. Additionally, half of the estimates diverge in no more than 9 percentage points from the real values. 
	\begin{algorithm}
		\caption{Coverage analysis}\label{alg:power test}
		\begin{algorithmic}[1] \footnotesize
			\State Simulate the dgp from Equation (\ref{eq:1})
			\State Estimate the model by drawing samples from the posterior distributions in subsection \ref{subsec:c} 
			\For{\texttt{$i=1,\dots,N$}} \For{\texttt{$t=1,\dots,T$}}
			\State $I_{it}=\mathbbm{1}\left[P_{it}\in \mathcal{P}_{\alpha}\right]$ where $\mathcal{P}_{\alpha}$ is the $1-\alpha$ highest density interval, that is, $\mathcal{P}_{\alpha}=\left\{Y_{it}\times\exp(\eta_i^+ + u_{it}^+):\pi(Y_{it}\times\exp(\eta_i^+ + u_{it}^+)|\bm\Theta,\bm Y, \bm X)\geq k(\alpha)\right\}$, $k(\alpha)$ is the largest constant such that $P(\mathcal{P}_{\alpha})\geq 1-\alpha$.  
			\EndFor
			\EndFor
			\State  Draw $S$ samples from $B(a,b)$ where $a=1+\sum_{i=1}^N\sum_{t=1}^T I_{it}$ and $b=1+N\times T-\sum_{i=1}^N\sum_{t=1}^T I_{it}$.  
			
			\State Obtain the point estimate of the coverage ($cov$) as the mean of these draws.
			\State Obtain the 95\% highest density interval from these draws.
		\end{algorithmic}
	\end{algorithm}
	\begin{table}[ht]
		\centering 
		\caption{Hidden population: Coverage results}\label{tab:MC}
		\begin{threeparttable}
			\begin{tabular}{lccc}
				\hline
				& Credibility level & Mean coverage & HDI \\ 
				\hline
				N=49 T=5 & 0.90 & 0.911 & (0.872,0.943) \\ 
				& 0.95 & 0.939 & (0.905,0.966) \\ 
				& 0.99 & 0.992 & (0.978,0.999) \\   \hline
				N=49 T=10 & 0.90 & 0.803 & (0.767,0.836) \\ 
				& 0.95 & 0.884 & (0.855,0.911) \\ 
				& 0.99 & 0.959 & (0.939,0.975) \\   \hline
				N=100 T=5 & 0.90 & 0.9 & (0.873,0.925) \\ 
				& 0.95 & 0.94 & (0.918,0.959) \\ 
				& 0.99 & 0.98 & (0.967,0.99) \\   \hline
				N=100 T=10 & 0.90 & 0.876 & (0.856,0.896) \\ 
				& 0.95 & 0.943 & (0.928,0.957) \\ 
				& 0.99 & 0.989 & (0.982,0.994) \\   \hline
				N=196 T=10 & 0.90 & 0.881 & (0.866,0.895) \\ 
				& 0.95 & 0.93 & (0.918,0.941) \\ 
				& 0.99 & 0.982 & (0.975,0.987) \\ 
				\hline
			\end{tabular}
			\begin{tablenotes}
				\footnotesize\textit{Note: Coverage analysis under different sample sizes and credibility levels. It suggests that the $1-\alpha$ highest density interval (HDI) for the hidden population, $\mathcal{P}_{\alpha}=\left\{Y_{it}\times\exp(\eta_i^+ + u_{it}^+):\pi(Y_{it}\times\exp(\eta_i^+ + u_{it}^+)|\bm\Theta,\bm Y, \bm X)\geq k(\alpha)\right\}$, $k(\alpha)$ is the largest constant such that $P(\mathcal{P}_{\alpha})\geq 1-\alpha$, has good coverage as they are very close to the nominal credibility levels.}
			\end{tablenotes}
		\end{threeparttable}
	\end{table}

	\begin{table}[ht]\caption{Summary statistics: mean absolute percentage error posterior estimates of hidden populations}
		\centering \label{tab:cov}
		\begin{threeparttable}
			\begin{tabular}{lcccc}
				\hline
				Sample Size & Average & Median &     \multicolumn{2}{c}{HDI} \\
				\hline
				N=49, T=5 & 0.2433 & 0.0886 & 0.0021 & 1.1326 \\ 
				N=49, T=10&  0.2055 & 0.0850 & 0.0018 & 0.7123 \\ 
				N=100, T=5 & 0.2109 & 0.0872 & 0.0004 & 0.7125 \\ 
				N=100, T=10 & 0.2195 & 0.0851 & 0.0013 & 0.8314 \\ 
				N=196, T=10 & 0.2418 & 0.0855 & 0.0008 & 0.9275 \\ 
				\hline
			\end{tabular}
			\begin{tablenotes}
				\item \footnotesize\textit{Note: Mean absolute percentage error (MAPE), $\text{MAPE}_{it}=\frac{1}{S}\sum_{s=1}^S\left|\frac{P_{it}-\widehat{\mathbb{E}}\left[P_{it}\right]}{P_{it}} \right|$ associated with the posterior predictive interval $\mathcal{P}_{\alpha}=\left\{Y_{it}\times\exp(\eta_i^+ + u_{it}^+):\pi(Y_{it}\times\exp(\eta_i^+ + u_{it}^+)|\bm\Theta,\bm Y, \bm X)\geq k(\alpha)\right\}$, $k(\alpha)$ is the largest constant such that $P(\mathcal{P}_{\alpha})\geq 1-\alpha$.}
			\end{tablenotes}
		\end{threeparttable}
	\end{table}
	
	\section{Crime suspect communities: the Medellín case}\label{Sec:App}
	Medellín is a natural experimental field to analyze crime. In particular, this city had a slightly increasing  homicide rate from the mid-1960s to 1980, then there is a remarkable increase between 1981 to 1991 passing from 52 to 388 homicides per one hundred thousand inhabitants \citep{Garcia2012}; this is mainly explained by the drug trafficking war of the local cartel against its competitors and the state. Then, there is a significant decrease of the homicide rate until 2008, reaching 46, that may be explained by the peace agreements with the guerrilla groups M-19 and EPL in 1990 and 1992, respectively, the dismantling of the local cartel, and death of its main figure in 1993, the intervention of the police and army (Orion operation) in \textit{Comuna 13}, a neighborhood characterized by no state law enforcement, and the demobilization of the paramilitary group \textit{Cacique Nutibara} in 2003, and other groups until 2006 \citep{Giraldo2015}. However, there is a sudden upturn in 2009, the homicide rate was 94, this may be explained by disputes to get control of the drug trafficking business after the extradition of former leaders. Since then, Medellín has experienced a steady decrease in the homicide rate achieving 20 in 2015, which is a historical low record for the city in the last 50 years.\par
	We got confidential information from Colombian police reports about residential address at capture moment of suspects associated with four criminal activities: homicide, drug dealing, motorcycle and car thefts (see Appendix \ref{Appendix}, Table \ref{tab:depdef}). We do not have the socioeconomic characteristics of these individuals. So, we calculate crime suspects rates per one hundred thousand inhabitants at the analytical region level in Medell\'in between 2011-2014.\footnote{However, we use rates per one million inhabitants for regressors in our models for coefficient scale purposes.} Then, as we consider environmental factors at a neighborhood level as potential drivers of ``crime communities'', we calculated for these analytical regions averages of socioeconomic characteristics that have been considered previously in crime literature \citep{Nunez2003,andresen2006spatial,hipp2007block,kakamu2008spatial,kikuchi2010neighborhood,arnio2012demography,He2015} using information from annual Living Standards Surveys (See Appendix \ref{Appendix}, Table \ref{tab:indepdef}). However, these surveys are not representative at a neighborhood level (325 units) in Medellín. Therefore, we use the \textit{max-p-region} algorithm \citep{duque2012max} obtaining 175 analytical regions which are more representative.
	This algorithm merges adjacent neighborhoods to create new analytical regions such that the algorithm minimizes within attribute heterogeneity, but maximizes this heterogeneity between the new analytical regions (see \cite{duque2012max} for details).\par
	
	Table \ref{tab:desc} in Appendix \ref{Appendix} shows descriptive statistics. The main take away from this table is the high level of heterogeneity regarding control variables between the analytical regions; this would suggest that Medellín is characterized by a very high level of social inequality. We also observe in this table a mean homicide rate equal to 49 per one hundred thousand inhabitants; there are
	many analytical regions without any homicide, whereas the central business district analytical region has a remarkable 2,062 rate, explained by relatively no many people living in this area. Observe that this is a common flaw when modeling crime rates. On the other hand, we model crime suspects residence per region where we also found that there are many analytical regions without any, but others with very high figures. Homicide suspects rate is the highest on average, followed by drug dealing, motorcycle thefts, and car thefts, respectively.
	\subsection{Unconditional analysis}
	We perform unconditional analysis to identify potential ``crime communities'' using standardized incidence ratios \citep{banerjee04}, $SIR_{it}=\frac{S_{it}}{E_{it}}$, where $S_{it}$ is number of crime suspects living in analytical region $i$ at time $t$ based on capture police reports, and $E_{it}=n_{it}\frac{\sum_{i=1}^N S_{it}}{\sum_{i=1}^N n_{it}}$ is the expected number of crime suspects, $n_{it}$ is the number of inhabitants in analytical region $i$ at time $t$.\par
	
	It is possible to find a $SIR$'s estimator by maximum likelihood; assuming $S_{it}|\eta_{it}$ distributes poisson, that is, $S_{it}|\eta_{it}\sim P(E_{it}{\eta}_{it})$. Then, the maximum likelihood (ML) estimator is $\hat{\eta}_{it} = SIR_{it}$. Nonetheless, assuming equi-dispersion may be non-realistic \citep{clayton87}. To get a more flexible model we assume that $S_{it}|\eta_{it}\sim P(E_{it}\eta_{it})$ such that $\eta_{it}$ distributes gamma, $\eta_{it} \sim G(\nu, \alpha)$; this implies that $\eta_{it}| S_{it}=s_{it}\sim G(s_{it} + \nu, E_{it} + \alpha)$. So, at the end, we get smoother ratios, through a prior distribution on $\eta_i$, overcoming equi-dispersion.\par 
	
	We are interested on the probability of a $SIR$ being higher than an observed value, that is, $H_0:S_{it}=E_{it}$ versus $H_1:S_{it}>E_{it}$. Then, if $S_{it}\sim P(E_{it}\eta_{it})$ under the null hypothesis $\eta_{it}=1$, $P(\eta_{it}> 1|s_{it},E_{it})$ is the posterior probability against the null hypothesis of evenly distributed rates across space, $P(\eta_{it}> 1|s_{it},E_{it})=1-P(\eta_{it}<1|s_{it},E_{it})=1-\int_0^{s_{it}}\frac{\eta_{it}^{s_{it}+\nu-1}\exp^{-\eta_i(E_{it} + \alpha)}(E_{it} + \alpha)^{s_{it} + \nu}}{\Gamma(s_{it} + \nu)}d\eta_{it}$ where $\Gamma(.)$ is the gamma function, and $\nu=\alpha=0.01$ to have non informative priors.\par
	
	Figure \ref{fig:1} shows probability maps, $P(\eta_{it}> 1|s_{it},E_{it})$ \citep{choynowski59}. This help to easily identify ``crime communities''. In particular, we observe that the central business district (map center) is a potential hot spot for homicide, drug dealing and car theft suspects. On the other hand, it seems that there are some other specialized ``crime communities''. Homicide suspects are located in the western (see Figure \ref{fig:1a}), drug dealing suspects at central-eastern (see Figure \ref{fig:1b}), car thefts at north-western (see Figure \ref{fig:1c}), and motorcycle thefts at north-eastern (see Figure \ref{fig:1d}).\par
	
	
	\begin{figure}[!h]
		\caption{Standardized incidence ratios: Medellín 2011-2014}
		\begin{subfigure}{0.475\textwidth}
			\centering
			\animategraphics[controls,width=\linewidth]{2}{homicidio}{}{}
			\caption{Suspects: homicide rates}
			\label{fig:1a}
		\end{subfigure}%
		\begin{subfigure}{0.475\textwidth}
			\centering
			\animategraphics[controls,width=\linewidth]{2}{estupefacientes}{}{}
			\caption{Suspects: drug dealing}
			\label{fig:1b}
		\end{subfigure}
		\begin{subfigure}{0.475\textwidth} 
			\centering
			\animategraphics[controls,width=\linewidth]{2}{autos}{}{}
			\caption{Suspects: car thefts}
			\label{fig:1c}
		\end{subfigure}%
		\begin{subfigure}{0.475\textwidth}
			\centering
			\animategraphics[controls,width=\linewidth]{2}{motos}{}{}
			\caption{Suspects: motorcycle thefts}
			\label{fig:1d}
		\end{subfigure}
		\label{fig:1}
		\caption*{\small \textit{{Note: Probability of rejecting the null hypothesis of evenly distributed rates of crime suspects in Medellín. This map identifies potential captured ``crime communities'' at 90\%, 95\% and 99\%.}}}
	\end{figure}
	\subsection{Conditional posterior results}
	
	Table \ref{tab:2} shows posterior estimates of our econometric proposal for four criminal activities: homicides, drug dealing, motorcycle and car thefts. Our dependent variable is $\log(1+Y_{it})\approx Y_{it}$ where $Y_{it}=\frac{S_{it}}{(n_{it}/100,000)}$.\par
	There are some interesting results regarding home location of crime suspects. It seems that homicide and drug dealing suspect communities are positively associated with high proportions of young males, and low population densities, but surrounded by neighborhoods with a high population density. This would suggest local urban displacement associated with these crime communities, although, this displacement seems not to be explicitly forced. Observe that these characteristics do not play any statistical significant role in motorcycle and car thefts suspect communities, which on the other hand, are positively associated with young unemployment rates. This suggests that local focused employment policies may reduce these criminal activities. Additionally, it seems that there is a kind of optimal location regarding these communities as they are located near middle income neighborhoods.\par 
	
	There are other specific statistical significant variables to each crime suspect community. For instance, homicide communities are positively associated with less immigrants, low male education and household expenditures, high household sizes and neighborhoods with a lower safety perception. Drug dealers communities are associated with less proportion of Caucasians, but higher levels of safety perception. The latter is also positively associated with motorcycle suspect communities, which in turn, is also positively associated with neighbors with high forced displacement and low safety perception. It seems that motorcycle thieves travel to close neighborhoods to commit their crimes (average travel time is 10 minutes from crime location to home location). Finally, car thieves communities is positively associated with a higher proportion of divorced males, more immigrants, less people per household locally and in surrounding neighborhoods.\par
	
	Table \ref{tab:3} reports posterior mean estimates of error components (one-sided and two-sided) associated with homicide, drug dealing, motorcycle and car thefts. We notice that $\lambda$ is approximately between 2 and 8, which implies a safe ground for inference in one-sided error models as shown from simulation exercises in Table \ref{tab:a1} and previous studies \citep{olson1980monte,simar2009inferences}. In addition, the percentage of total variability due to spatial effects is between 20\% (homicides) and 13\% (car thefts), where $\frac{\sigma_v}{0.7\left(\sum_{i\sim j}w_{ij}\right)^{Ave}}$ is the marginal standard deviation due to spatial effects \citep{ramirez2017welfare}, $\left(\sum_{i\sim j}w_{ij}\right)^{Ave}$ is the average number of neighbors. This highlights the relevance of this effect. Finally, the posterior mean estimate of permanent percentage of potential covered (uncaptured) crime suspects ($\hat{\mathbb{E}}(1-\exp(\eta_i^+))$) fluctuates between 18\% (car thefts) and 26\% (drug dealing), and the total (permanent plus transient, $\hat{\mathbb{E}}(1-\exp(\eta_i^+ + u_{it}^+))$) is between 32\% (car thefts) and 57\% (homicides and drug dealing). This means that on average the highest transient effect were associated with homicides (33\%).\par
	
	However, the former figures have a lot of heterogeneity through time and space. Figure \ref{fig:3} shows analytical regions specific total percentages (permanent and transient) of potentially still covered suspects by crime and time. Regarding homicide (top-left panel), it seems that between 2011 and 2013 there was a hot spot of potential covered crime communities in the most western area (analytical region 121) with percentages over 90\%. However, this situation drastically changed in 2014 for this area, it seems that these hot spots moved a little bit to east (analytical regions 121 and 177). In addition, there is a cluster of homicide crime communities from the central east (analytical region 80) to the north limit of the central business center (analytical region 163) for this last year.
	
	Drug dealing have similar pattern to homicides regarding the hot spot in the most western area between 2011 and 2013 (top-right panel). However, analytical region 121 still seems to be a drug dealers community in 2014. Observe that similar patters regarding statistically relevant variables was also found in estimation results, and coincides with the violent history of Medell\'in due to the drug trafficking war of gangs for business control. Both crime activities (homicide and drug dealing) have uncaptured percentage rates as high as 90\%.
	
	Car thefts communities are located in the central-north area near the west riverside of the Medell\'in river (bottom-left panel). This river is a geographical barrier between the west and the east of the city, and plays an important role regarding crime communities. It seems that there is a hot spot composed by analytical regions 44 to 47. Another hot spot is composed by analytical regions 95, 71 and 73 located on the central-west. It seems that this is also a community of motorcycle thieves. Observe that the percentage of uncaptured car thieves is as high as 50\%, whereas this figure is as high as 70\% for motorcycle thieves.
	
	\begin{table}[ht]\caption{Bayesian posterior estimates: capture rates of crime suspects}\label{tab:2}
		\begin{threeparttable}
			\centering \resizebox{\textwidth}{!}{%
				\begin{tabular}{lcccc}
					\hline
					& Homicides & Drug dealing & Motorcycle thefts & Car thefts \\  
					\hline
					Divorced males 	&	0.003	&	-0.056	&	-0.0596	&	 0.0861** \\ 
					Caucasian 	&	-0.0174	&	 -0.0753** 	&	8.00E-04	&	 -0.007 \\ 
					Immigrants 	&	 -0.2675*** 	&	-0.1181	&	-0.0348	&	 0.162*** \\ 
					$\text{Unemployment 24-15}$ 	&	0.0969	&	0.0305	&	 0.0835** 	&	 0.0553* \\ 
					Male population 24-15 	&	 0.7973*** 	&	 0.3704** 	&	-0.0862	&	 0.0722 \\ 
					Male education 	&	 -0.0649* 	&	-0.0271	&	0.0326	&	 -0.0309 \\ 
					Illiteracy 	&	0.0379	&	0.0345	&	-0.0199	&	 0.0524* \\ 
					Forced displacement 	&	0.0201	&	0.0411	&	-0.0304	&	 0.0131 \\ 
					Safety perception 	&	-0.0676	&	 0.2083* 	&	 0.1952** 	&	 -0.0242 \\ 
					Expenditure per capita 	&	 -0.1493** 	&	-0.0261	&	0.0203	&	 -0.0185 \\ 
					Population density 	&	 -0.0135*** 	&	 -0.0113*** 	&	0.0029	&	 -0.0024 \\ 
					People per household 	&	 0.1123** 	&	0.0391	&	0.0562	&	 -0.0763** \\ 
					Middle income 	&	-0.1624	&	-0.2189	&	-0.0676	&	 -0.0864 \\ 
					High income 	&	0.2064	&	-0.3982	&	-0.2071	&	 -0.1042 \\ 
					Divorced males (spatial lag)	&	0.022	&	0.0215	&	-0.0051	&	 -0.0041 \\ 
					Caucasian (spatial lag)	&	-0.0152	&	0.0131	&	0.0127	&	 -3e-04 \\ 
					Immigrants (spatial lag)	&	0.0506	&	0.0121	&	0.0286	&	 -0.018 \\ 
					$\text{Unemployment 24-15}$  (spatial lag) 	&	0.026	&	-0.0078	&	0.011	&	 -0.0016 \\ 
					Male population 24-15 (spatial lag)	&	-0.0309	&	-0.0196	&	0.0168	&	 -0.0219 \\ 
					Male education (spatial lag)	&	0.025	&	0.0096	&	-0.018	&	 -0.0083 \\ 
					Illiteracy (spatial lag)	&	-0.0207	&	-8.00E-04	&	-0.0036	&	 -0.0128 \\ 
					Forced displacement (spatial lag)	&	0.0106	&	-0.0046	&	 0.0186** 	&	 0.0036 \\ 
					Safety perception (spatial lag)	&	 -0.085* 	&	-0.0417	&	 -0.0634** 	&	 0.0199 \\ 
					Expenditure per capita (spatial lag)	&	0.0209	&	0.0224	&	-0.0206	&	 -0.01 \\ 
					Population density (spatial lag)	&	 0.0043*** 	&	 0.0022** 	&	8.00E-04	&	 -3e-04 \\ 
					People per household (spatial lag)	&	0.0274	&	0.0144	&	0.0147	&	 0.0236* \\ 
					Middle income (spatial lag)	&	0.0436	&	0.0473	&	 0.0905** 	&	 0.065* \\ 
					High income (spatial lag)	&	-0.0123	&	-0.0849	&	0.1249	&	 0.0574 \\ 
					
					\hline
			\end{tabular}}
			\begin{tablenotes}
				\item \footnotesize \textit{Notes: Posterior mean estimates. ***, ** and * are statistically significant variables at 1\%, 5\% and 10\%.}
				\item \footnotesize \textit{In particular, 99\%, 95\% and 90\% highest density intervals do not embraces zero.}
			\end{tablenotes}
		\end{threeparttable}
	\end{table}
	
	\begin{table}[ht] \caption{Posterior estimates: error components}\label{tab:3}
		\begin{threeparttable}
			\begin{tabular}{lcccc}
				\hline
				& Homicides & Drug dealing & Motorcycle thefts & Car thefts  \\ 
				\hline
				$\hat{\mathbb{E}}\big[$1-exp(-$\eta_i^+)\big]$ & 0.24 & 0.26 & 0.21 & 0.18 \\ 
				$\sigma_\eta$ & 0.35 & 0.40 & 0.31 & 0.27 \\ 
				$\hat{\mathbb{E}}\big[$1-exp(-($\eta_i^++u_{it}^+))\big]$ & 0.57 & 0.57 & 0.42 & 0.32 \\ 
				$\sigma_u$  & 0.81 & 0.75 & 0.41 & 0.24 \\ 
				$\sigma_v$  & 0.73 & 0.55 & 0.43 & 0.24 \\ 
				$\sigma_\alpha$  & 0.11 & 0.17 & 0.10 & 0.07 \\ 
				$\sigma_\epsilon$  & 0.17 & 0.14 & 0.25 & 0.26 \\ 
				$\lambda=\frac{\sigma_{\eta}+\sigma_u}{\sigma_{\epsilon}}$ & 6.77 & 8.21 & 2.91 & 1.94 \\ 
				\hline
			\end{tabular}
			\begin{tablenotes}
				\item \small \textit{Notes: Posterior mean estimates. Standard deviations stochastic components (two-sided and one-sided), permanent and total percentage of potential still covered crime suspects, and total one-sided variation to stochastic error variation ratio.}
			\end{tablenotes}
		\end{threeparttable}
	\end{table}
	
	\begin{figure}[!h]
		\caption{Percentage of potential uncaptured crime suspects: Medellín 2011-2014}
		\begin{subfigure}{0.475\textwidth}
			\centering
			\animategraphics[controls,width=\linewidth]{2}{homicides}{}{}
			\caption{Suspects: homicide rates}
			\label{fig:3a}
		\end{subfigure}%
		\begin{subfigure}{0.475\textwidth}
			\centering
			\animategraphics[controls,width=\linewidth]{2}{drug_dealing}{}{}
			\caption{Suspects: drug dealing}
			\label{fig:3b}
		\end{subfigure}
		\begin{subfigure}{0.475\textwidth} 
			\centering
			\animategraphics[controls,width=\linewidth]{2}{automobile_thefts}{}{}
			\caption{Suspects: car thefts}
			\label{fig:3c}
		\end{subfigure}%
		\begin{subfigure}{0.475\textwidth}
			\centering
			\animategraphics[controls,width=\linewidth]{2}{motorcycle_thefts}{}{}
			\caption{Suspects: motorcycle thefts}
			\label{fig:3d}
		\end{subfigure}
		\label{fig:3}
		\caption*{\small \textit{Note: Percentage of uncaptured crime suspects estimates, $\hat{\mathbb{E}}\big[$1-exp(-($\eta_i^++u_{it}^+))\big]$. It seems that homicide and drug dealing have similar hot spots located at the most west analytical region (top panels), whereas car and motorcycle have a hot spot at the central-west (bottom panels).}}
	\end{figure}
	
	\section{Concluding remarks}\label{Sec:Conc}
	
	We propose a Bayesian approach to perform inference regarding ``hidden populations'' at analytical region level such as criminal activity. We extend a generalized random effects model including spatial effects where ``hidden populations'' are taken into account using one-sided errors. Simulation exercises suggest that our proposal has good sampling properties regarding point estimates and ``hidden population'' predictions.
	
	Our application based on home place of crime suspects suggest that there is association between homicide and drug dealing which has caused potential urban displacement to neighbourhoods near these crime communities. This is also supported by historical facts and higher levels of uncaptured crime suspects, which are as high as 90\% in the hot spots of both activities. On the other hand, motorcycle and car thefts have lower uncaptured rates (70\% and 50\%, respectively), and both activities are associated with high local unemployment rates, which would suggest that focused employment policies would mitigate these activities.
	
	Due to the elapsed time between crime moment and reported time using CCTV monitoring, we suggest that a potentially good strategy to capture crime suspects is to lock down the potential destination neighborhood of criminals. Our modelling strategy would help to predict this potential destination neighborhoods as we identified potential ``crime communities''. On the other hand, focused policy interventions targeting these communities with specific education and employment objectives would help to structurally reduce crime activity.   
	
	Future research should take into account sensitivity of our proposal to spatial contiguity criteria, where contiguity matrices can be selected based on Bayes factors or performing Bayesian model average to take into account this uncertainty source.
	
	\clearpage \bibliography{cites.bib} \newpage

\begin{thebibliography}{}

\bibitem[Aigner et~al., 1977]{Aigner1977}
Aigner, D., Lovell, K., and Schmidt, P. (1977).
\newblock Formulation and estimation of stochastic production function models.
\newblock {\em Journal of Econometrics}, 6:21--37.

\bibitem[Andresen, 2006]{andresen2006spatial}
Andresen, M.~A. (2006).
\newblock A spatial analysis of crime in vancouver, british columbia: A
  synthesis of social disorganization and routine activity theory.
\newblock {\em The Canadian Geographer/Le G{\'e}ographe canadien},
  50(4):487--502.

\bibitem[Arnio and Baumer, 2012]{arnio2012demography}
Arnio, A.~N. and Baumer, E.~P. (2012).
\newblock Demography, foreclosure, and crime: Assessing spatial heterogeneity
  in contemporary models of neighborhood crime rates.
\newblock {\em Demographic Research}, 26:449--486.

\bibitem[Banerjee et~al., 2004]{banerjee04}
Banerjee, S., Carlin, B., and Gelfand, A. (2004).
\newblock {\em Hierarchical Modeling and Analysis for Spatial Data}.
\newblock Chapman \& {H}all/{CRC}.

\bibitem[Banerjee et~al., 2014]{banerjee2014hierarchical}
Banerjee, S., Carlin, B.~P., and Gelfand, A.~E. (2014).
\newblock {\em Hierarchical modeling and analysis for spatial data}.
\newblock Crc Press.

\bibitem[Besag, 1991]{besag1991}
Besag, J. (1991).
\newblock Bayesian image restoration, with two applications in spatial
  statistics.
\newblock {\em Annals of the institute of statistical mathematics},
  43(1):1--20.

\bibitem[Bourguignon et~al., 2003]{Nunez2003}
Bourguignon, F., Nuñez, J., and Sanchez, F. (2003).
\newblock A structural model of crime and inequality in {C}olombia.
\newblock {\em Journal of the European Economic Association}, 1(2-3):440--449.

\bibitem[Choynowski, 1959]{choynowski59}
Choynowski, M. (1959).
\newblock Maps based on probabilities.
\newblock {\em Journal of the American Statistical Association},
  54(286):385--388.

\bibitem[Clayton and Kaldor, 1987]{clayton87}
Clayton, D. and Kaldor, J. (1987).
\newblock Empirical bayes estimates of age-standardized relative risks for use
  in disease mapping.
\newblock {\em Biometrics}, 43(3):671--681.

\bibitem[Collazos et~al., 2020]{Collazos2020}
Collazos, D., García, E., Mejía, D., Ortega, D., and Tobón, S. (2020).
\newblock Hot spots policing in a high-crime environment: an experimental
  evaluation in medellín.
\newblock {\em Journal of Experimental Criminology}, Forthcoming.

\bibitem[Dom\'inguez-Molina et~al., 2003]{Dominguez2003}
Dom\'inguez-Molina, J., Gonz\'alez-Far\'ias, G., and Ramos-Quiroga, R. (2003).
\newblock Skew-normality in stochastic frontier analysis.
\newblock Technical report, CIMAT, Mexico.

\bibitem[Druska and Horrace, 2003]{druska2003generalized}
Druska, V. and Horrace, W.~C. (2003).
\newblock {\em Generalized moments estimation for spatial panel data}.
\newblock National Bureau of Economic Research.

\bibitem[Duque et~al., 2012]{duque2012max}
Duque, J.~C., Anselin, L., and Rey, S.~J. (2012).
\newblock The max-p-regions problem.
\newblock {\em Journal of Regional Science}, 52(3):397--419.

\bibitem[Eaton, 1983]{eaton1983multivariate}
Eaton, M.~L. (1983).
\newblock Multivariate statistics: a vector space approach.
\newblock {\em John Wiley \& Sons, INC., 605 Third Abe., New York, NY 10158,
  USA, 1983, 512}.

\bibitem[Garc\'ia et~al., 2012]{Garcia2012}
Garc\'ia, H.~I., Giraldo, C.~A., L\'opez, M.~V., del Pilar~Pastor, M., Cardona,
  M., Tapias, C.~E., Cuartas, D., G\'omez, V., and Vera, C.~Y. (2012).
\newblock Thirty years of homicides in medell\'n, colombia, 1979-2008.
\newblock {\em Cad. Sa\'ude P\'ublica}, 28(9):1699--1712.

\bibitem[Giraldo-Ram\'irez and Preciado-Restrepo, 2015]{Giraldo2015}
Giraldo-Ram\'irez, J. and Preciado-Restrepo, A. (2015).
\newblock Medellín, from theater of war to security laboratory.
\newblock {\em Stability: International Journal of Security and Development},
  4(1):1--14.

\bibitem[Glass et~al., 2016]{glass2016spatial}
Glass, A.~J., Kenjegalieva, K., and Sickles, R.~C. (2016).
\newblock A spatial autoregressive stochastic frontier model for panel data
  with asymmetric efficiency spillovers.
\newblock {\em Journal of Econometrics}, 190(2):289--300.

\bibitem[Gude et~al., 2018]{gude2018heterogeneous}
Gude, A., {\'A}lvarez, I., and Orea, L. (2018).
\newblock Heterogeneous spillovers among spanish provinces: a generalized
  spatial stochastic frontier model.
\newblock {\em Journal of Productivity Analysis}, 50(3):155--173.

\bibitem[Hipp, 2007]{hipp2007block}
Hipp, J.~R. (2007).
\newblock Block, tract, and levels of aggregation: Neighborhood structure and
  crime and disorder as a case in point.
\newblock {\em American Sociological Review}, 72(5):659--680.

\bibitem[Horrace and Wright, 2019]{horrace2019stationary}
Horrace, W.~C. and Wright, I.~A. (2019).
\newblock Stationary points for parametric stochastic frontier models.
\newblock {\em Journal of Business \& Economic Statistics}, pages 1--26.

\bibitem[Kakamu et~al., 2008]{kakamu2008spatial}
Kakamu, K., Polasek, W., and Wago, H. (2008).
\newblock Spatial interaction of crime incidents in japan.
\newblock {\em Mathematics and Computers in Simulation}, 78(2-3):276--282.

\bibitem[Kikuchi, 2010]{kikuchi2010neighborhood}
Kikuchi, G. (2010).
\newblock {\em Neighborhood structures and crime: a spatial analysis}.
\newblock LFB Scholarly Pub. LLC.

\bibitem[Kirk, 2006]{Kirk2006}
Kirk, D.~S. (2006).
\newblock Examining the divergence across self-report and official data sources
  on inferences about the adolescent life-course of crime.
\newblock {\em J Quant Criminol}, 22:107--129.

\bibitem[Li et~al., 2015]{He2015}
Li, H., Antonio, P., Desheng, L., and Shiguo, J. (2015).
\newblock Temporal stability of model parameters in crime rate analysis: An
  empirical examination.
\newblock {\em Applied Geography}, 58:141--152.

\bibitem[Makie{\l}a, 2017]{makiela2017bayesian}
Makie{\l}a, K. (2017).
\newblock Bayesian inference and gibbs sampling in generalized true
  random-effects models.
\newblock {\em Central European Journal of Economic Modelling and
  Econometrics}, pages 69--95.

\bibitem[Mastromarco et~al., 2016]{mastromarco2016modelling}
Mastromarco, C., Serlenga, L., and Shin, Y. (2016).
\newblock Modelling technical efficiency in cross sectionally dependent
  stochastic frontier panels.
\newblock {\em Journal of Applied Econometrics}, 31(1):281--297.

\bibitem[Meeusen and van Den~Broeck, 1977]{Meeusen1977}
Meeusen, W. and van Den~Broeck, J. (1977).
\newblock Efficiency estimation from {C}obb-{D}ouglas production functions with
  composed error.
\newblock {\em International Economic Review}, 18(2):435--444.

\bibitem[Olson et~al., 1980]{olson1980monte}
Olson, J.~A., Schmidt, P., and Waldman, D.~M. (1980).
\newblock A monte carlo study of estimators of stochastic frontier production
  functions.
\newblock {\em Journal of Econometrics}, 13(1):67--82.

\bibitem[Ram{\'\i}rez~Hassan and Montoya~Bland{\'o}n, 2017]{ramirez2017welfare}
Ram{\'\i}rez~Hassan, A. and Montoya~Bland{\'o}n, S. (2017).
\newblock Welfare gains of the poor: An endogenous bayesian approach with
  spatial random effects.
\newblock {\em Econometric Reviews}, pages 1--18.

\bibitem[Schmidt et~al., 2009]{schmidt2009spatial}
Schmidt, A.~M., Moreira, A.~R., Helfand, S.~M., and Fonseca, T.~C. (2009).
\newblock Spatial stochastic frontier models: accounting for unobserved local
  determinants of inefficiency.
\newblock {\em Journal of Productivity Analysis}, 31(2):101--112.

\bibitem[Simar and Wilson, 2009]{simar2009inferences}
Simar, L. and Wilson, P.~W. (2009).
\newblock Inferences from cross-sectional, stochastic frontier models.
\newblock {\em Econometric Reviews}, 29(1):62--98.

\bibitem[Sun et~al., 1999]{sun1999posterior}
Sun, D., Tsutakawa, R.~K., and Speckman, P.~L. (1999).
\newblock Posterior distribution of hierarchical models using car (1)
  distributions.
\newblock {\em Biometrika}, 86(2):341--350.

\bibitem[Sánchez-González et~al., 2020]{Sanchez2020}
Sánchez-González, J., Restrepo, D., and Ramírez-Hassan, A. (2020).
\newblock Inefficiency and bank failure: A joint bayesian estimation method of
  stochastic frontier and hazards models.
\newblock {\em Economic Modelling}, Forthcoming.

\bibitem[Tanner and Wong, 1987]{Tanner1987}
Tanner, M.~A. and Wong, W.~H. (1987).
\newblock The calculation of posterior distributions by data augmentation.
\newblock {\em Journal of the American Statistical Association},
  82(398):528--540.

\bibitem[Tsionas and Kumbhakar, 2014]{tsionas2014firm}
Tsionas, E.~G. and Kumbhakar, S.~C. (2014).
\newblock Firm heterogeneity, persistent and transient technical inefficiency:
  A generalized true random-effects model.
\newblock {\em Journal of Applied Econometrics}, 29(1):110--132.

\bibitem[Tsionas and Michaelides, 2016]{tsionas2016spatial}
Tsionas, E.~G. and Michaelides, P.~G. (2016).
\newblock A spatial stochastic frontier model with spillovers: Evidence for
  italian regions.
\newblock {\em Scottish Journal of Political Economy}, 63(3):243--257.

\bibitem[van~der Heijden~P. and Cruyff~M., 2014]{Heijden2014}
van~der Heijden~P. and Cruyff~M., B.~D. (2014).
\newblock Capture recapture to estimate criminal populations.
\newblock In G., B. and D., W., editors, {\em Encyclopedia of Criminology and
  Criminal Justice}, pages 267--276. Springer, New York, NY.

\bibitem[Waldman, 1982]{waldman1982stationary}
Waldman, D.~M. (1982).
\newblock A stationary point for the stochastic frontier likelihood.
\newblock {\em Journal of Econometrics}, 18(2):275--279.

\bibitem[Wang and Schmidt, 2002]{Wang2002}
Wang, H.-J. and Schmidt, P. (2002).
\newblock One-step and two-step estimation of the effects of exogenous
  variables on technical efficiency levels.
\newblock {\em Journal of Productivity Analysis}, 18(2):129--144.

\end{thebibliography}
	\appendix
	\section{Appendix}\label{Appendix}
	\subsection{Descriptive statistics}\label{Subsec:desc_stats}
	
	\begin{table}[ht]
		\caption{Definition: dependent variables}\label{tab:depdef}
		\centering
		\begin{tabular}{rll}
			\hline
			Variable & Description \\ 
			\hline
			Homicide & \begin{minipage}[t]{0.75\columnwidth}%
				Amount of captured homicide suspects living at analytical region $i$ in year $t$ per one hundred thousand inhabitants.%
			\end{minipage} \\ 
			Drug dealing & \begin{minipage}[t]{0.75\columnwidth}%
				Amount of captured drug dealing suspects living at analytical region $i$ in year $t$ per one hundred thousand inhabitants.%
			\end{minipage}	\\ 
			Car theft & \begin{minipage}[t]{0.75\columnwidth}%
				Amount of captured car theft suspects living at analytical region $i$ in year $t$ per one hundred thousand inhabitants.%
			\end{minipage} \\ 
			Motorcycles theft & \begin{minipage}[t]{0.75\columnwidth}%
				Amount of captured motorcycle theft suspects living at analytical region $i$ in year $t$ per one hundred thousand inhabitants.%
			\end{minipage} \\
			\hline
		\end{tabular}
		\begin{tablenotes}
			\item \small \textit{Note: This information comes from police reports between 2011 and 2014.}
		\end{tablenotes}
	\end{table}

	\begin{table}[ht]
		\caption{Definition: control variables}\label{tab:indepdef}
		\centering
		\begin{tabular}{rll}
			\hline
			Variable & Description \\ 
			\hline
			Male population 15-24 & \begin{minipage}[t]{0.7\columnwidth}%
				Proportion of males between 15 and 24 years old
			\end{minipage} \\ 
			Divorced males & Proportion of divorced males\\ 
			Caucasian & Proportion of white race people\\ 
			Immigrants & Proportion of people who migrated to region\\ 
			Unemployment 15-24 & \begin{minipage}[t]{0.7\columnwidth}%
				Proportion of unemployed people between 15 and 24 years old
			\end{minipage} \\
			Low income & Proportion of low income households\\ 
			Middle income & Proportion of middle income households\\ 
			High income & Proportion of high income households\\ Male education & \begin{minipage}[t]{0.7\columnwidth}%
				Proportion of males head of household with college degree or higher%
			\end{minipage}\\ 
			Illiteracy & \begin{minipage}[t]{0.7\columnwidth}%
				Proportion of population older than 15 years old that does not know how to read or write%
			\end{minipage} \\ 
			Forced displacement & Proportion of families that suffer forced displacement\\ 
			People per household & Average number of people per household\\ 
			Population density & Average number of people per $km^2$\\ 
			Safety perception & Proportion of population who feels unsafe\\ 
			Expenditure per capita & Average per capita expenditure in COP\$\\ 
			\hline
		\end{tabular}
		\begin{tablenotes}
			\item \small \textit{Notes: Proportion variables are measured in terms of total per million inhabitants in analytical regions. Information comes from Medell\'in living standards survey between 2011 and 2014.}
		\end{tablenotes}
	\end{table}
	
	\begin{landscape}
		\begin{table}[ht]\caption{Descriptive statistics: analytical regions in Medellín, 2011-2014.}\label{tab:desc}
			\begin{threeparttable}
				\begin{tabular}{rrrrrrr}
					& Mean & Overall s.d. & Between s.d. & Within s.d. & Min. & Max. \\ 
					\hline
					Male population 24-15 & 8.81 & 2.38 & 1.61 & 1.76 & 1.75 & 17.00 \\ 
					Divorced males & 1.54 & 0.99 & 0.59 & 0.80 & 0.00 & 5.91 \\ 
					Caucasian & 20.16 & 10.89 & 6.80 & 8.52 & 0.00 & 69.66 \\ 
					Immigrants & 46.81 & 14.12 & 11.94 & 7.58 & 11.76 & 86.29 \\ 
					$\text{Unemployment 24-15}^+$ & 1.33 & 1.08 & 0.78 & 0.75 & 0.00 & 5.50 \\ 
					Male education & 9.11 & 12.27 & 11.89 & 3.13 & 0.00 & 61.22 \\ 
					Illiteracy & 2.06 & 1.79 & 1.28 & 1.25 & 0.00 & 10.06 \\ 
					Forced displacement & 4.89 & 5.40 & 4.58 & 2.88 & 0.00 & 39.02 \\ 
					Safety perception & 81.74 & 13.83 & 11.02 & 8.38 & 18.03 & 100.00 \\ 
					Expenditure per capita & 1,390,812 & 839,837 & 807,886 & 235,473 & 455,313 & 6,179,126 \\ 
					Population density & 29.44 & 14.90 & 14.92 & 0.74 & 0.12 & 83.58 \\ 
					People per household & 3.65 & 0.50 & 0.40 & 0.30 & 2.00 & 5.00 \\ 
					Low income & 43.36 & 43.34 & 43.42 & 1.17 & 0.00 & 100.00 \\ 
					Middle income & 44.28 & 40.84 & 40.90 & 1.38 & 0.00 & 100.00 \\ 
					High income & 12.36 & 28.42 & 28.47 & 0.92 & 0.00 & 100.00 \\ 
					Homicide rate & 49.30 & 125.88 & 116.58 & 48.11 & 0.00 & 2,062.07 \\ 
					Suspects homicide rate & 9.11 & 18.17 & 12.18 & 13.51 & 0.00 & 309.31 \\ 
					Suspects drug dealing rate & 7.37 & 13.04 & 9.44 & 9.02 & 0.00 & 103.10 \\ 
					Suspects motorcycle thefts rate & 4.92 & 9.68 & 7.14 & 6.55 & 0.00 & 77.33 \\ 
					Suspects car thefts rate & 2.33 & 6.09 & 3.42 & 5.04 & 0.00 & 50.53 \\ 
					\hline
				\end{tabular}
				\begin{tablenotes}
					\item \small \textit{Notes: $^+$ this are pseudo employment measures, since we computed them by using total population and not the economically active population. Proportion variables are measured in terms of total per million inhabitants in analytical regions. Information comes from Medell\'in living standards survey between 2011 and 2014.}\\
				\end{tablenotes}
			\end{threeparttable}
		\end{table}
	\end{landscape}
	\subsection{Robustness checks}\label{Subsec:Rob}
	\begin{landscape}
		\begin{table}[p] \caption{Sampling properties of Bayes estimators: Varying $\lambda=\frac{\sigma_{u^+}+\sigma_{\eta^+}}{\sigma_{\epsilon}}$}\label{tab:a1}
			\centering  \resizebox{1.4\textheight}{!}{%
				\begin{tabular}{ccccccccccccccccccccccccccccccc}
					\hline    & $\lambda$ &
					\multicolumn{3}{c}{$\eta^+$} &
					\multicolumn{3}{c}{$\sigma_\eta$}&
					\multicolumn{3}{c}{$u^+$}&
					\multicolumn{3}{c}{$\sigma_u$} &
					\multicolumn{2}{c}{$v$}&
					\multicolumn{3}{c}{$\sigma_v$} &
					\multicolumn{3}{c}{$\sigma_\alpha$} &
					\multicolumn{3}{c}{$\sigma_\epsilon$} &
					\multicolumn{3}{c}{$\beta_1$} &
					\multicolumn{3}{c}{$\beta_2$} \\
					\cmidrule(r){3-5}  \cmidrule(r){6-8}  \cmidrule(r){9-11}  \cmidrule(r){12-14}  \cmidrule(r){15-16}  \cmidrule(r){17-19}  \cmidrule(r){20-22}  \cmidrule(r){23-25}  \cmidrule(r){26-28}  \cmidrule(r){29-31}
					& Mean &
					Mean &
					Median &
					$\hat{\E}(\rho)$    & Mean &
					\multicolumn{2}{c}{HDI}  &   Mean &
					Median &
					$\hat{\E}(\rho)$    & Mean &
					\multicolumn{2}{c}{HDI} & 
					Median &
					$\hat{\E}(\rho)$    & Mean &
					\multicolumn{2}{c}{HDI} & Mean &
					\multicolumn{2}{c}{HDI} & Mean &
					\multicolumn{2}{c}{HDI} & Mean &
					\multicolumn{2}{c}{HDI}& Mean &
					\multicolumn{2}{c}{HDI}  \\ 
					\hline
					True & 0.100 & -0.064 & -0.063 &  & 0.071 &  &  & -0.024 & -0.022 &  & 0.029 &  &  & 0.050 &  & 0.400 &  &  & 0.100 &  &  & 1.000 &  &  & 0.500 &  &  & -0.500 &  &  \\ 
					Estimated & 0.555 & -0.212 & -0.195 & -0.006 & 0.330 & 0.250 & 0.408 & -0.133 & -0.131 & -0.004 & 0.173 & 0.125 & 0.220 & 0.022 & 0.337 & 0.649 & 0.219 & 0.999 & 0.141 & 0.003 & 0.342 & 0.906 & 0.826 & 0.988 & 0.396 & 0.277 & 0.538 & -0.521 & -0.570 & -0.467 \\ 
					N=49, T=5 &  &  &  &  &  &  &  &  &  &  &  &  &  &  &  &  &  &  &  &  &  &  &  &  &  &  &  &  &  &  \\ 
					True & 0.100 & -0.060 & -0.054 &  & 0.071 &  &  & -0.023 & -0.019 &  & 0.029 &  &  & 0.013 &  & 0.400 &  &  & 0.100 &  &  & 1.000 &  &  & 0.500 &  &  & -0.500 &  &  \\ 
					Estimated & 0.336 & -0.142 & -0.139 & 0.058 & 0.210 & 0.179 & 0.242 & -0.094 & -0.093 & -0.002 & 0.119 & 0.096 & 0.140 & 0.006 & 0.612 & 0.567 & 0.432 & 0.697 & 0.119 & 0.004 & 0.229 & 0.980 & 0.953 & 1.014 & 0.461 & 0.415 & 0.502 & -0.499 & -0.515 & -0.483 \\ 
					N=196, T=10 &  &  &  &  &  &  &  &  &  &  &  &  &  &  &  &  &  &  &  &  &  &  &  &  &  &  &  &  &  &  \\ 
					True & 0.223 & -0.142 & -0.140 &  & 0.159 &  &  & -0.054 & -0.048 &  & 0.064 &  &  & 0.050 &  & 0.400 &  &  & 0.100 &  &  & 1.000 &  &  & 0.500 &  &  & -0.500 &  &  \\ 
					Estimated & 0.590 & -0.232 & -0.211 & 0.029 & 0.348 & 0.257 & 0.430 & -0.138 & -0.138 & -0.003 & 0.179 & 0.132 & 0.229 & 0.025 & 0.366 & 0.660 & 0.263 & 1.031 & 0.069 & 0.003 & 0.170 & 0.892 & 0.816 & 0.960 & 0.400 & 0.275 & 0.535 & -0.514 & -0.564 & -0.463 \\ 
					N=49, T=5 &  &  &  &  &  &  &  &  &  &  &  &  &  &  &  &  &  &  &  &  &  &  &  &  &  &  &  &  &  &  \\ 
					True & 0.223 & -0.135 & -0.121 &  & 0.159 &  &  & -0.051 & -0.043 &  & 0.064 &  &  & 0.013 &  & 0.400 &  &  & 0.100 &  &  & 1.000 &  &  & 0.500 &  &  & -0.500 &  &  \\ 
					Estimated & 0.375 & -0.164 & -0.158 & 0.104 & 0.233 & 0.199 & 0.268 & -0.103 & -0.103 & -0.000 & 0.131 & 0.107 & 0.155 & -0.003 & 0.617 & 0.590 & 0.448 & 0.716 & 0.040 & 0.003 & 0.095 & 0.970 & 0.942 & 1.003 & 0.463 & 0.418 & 0.507 & -0.499 & -0.515 & -0.483 \\ 
					N=196, T=10 &  &  &  &  &  &  &  &  &  &  &  &  &  &  &  &  &  &  &  &  &  &  &  &  &  &  &  &  &  &  \\ 
					True & 0.316 & -0.201 & -0.198 &  & 0.226 &  &  & -0.076 & -0.068 &  & 0.090 &  &  & 0.050 &  & 0.400 &  &  & 0.100 &  &  & 1.000 &  &  & 0.500 &  &  & -0.500 &  &  \\ 
					Estimated & 0.631 & -0.255 & -0.231 & 0.053 & 0.368 & 0.272 & 0.457 & -0.149 & -0.147 & 0.004 & 0.192 & 0.135 & 0.249 & 0.035 & 0.379 & 0.666 & 0.258 & 1.053 & 0.054 & 0.002 & 0.125 & 0.887 & 0.820 & 0.963 & 0.404 & 0.284 & 0.527 & -0.509 & -0.558 & -0.461 \\ 
					N=49, T=5 &  &  &  &  &  &  &  &  &  &  &  &  &  &  &  &  &  &  &  &  &  &  &  &  &  &  &  &  &  &  \\ 
					True & 0.316 & -0.191 & -0.172 &  & 0.226 &  &  & -0.072 & -0.061 &  & 0.090 &  &  & 0.013 &  & 0.400 &  &  & 0.100 &  &  & 1.000 &  &  & 0.500 &  &  & -0.500 &  &  \\ 
					Estimated & 0.419 & -0.187 & -0.179 & 0.144 & 0.257 & 0.215 & 0.297 & -0.118 & -0.118 & 0.001 & 0.149 & 0.119 & 0.180 & -0.004 & 0.608 & 0.610 & 0.462 & 0.752 & 0.030 & 0.003 & 0.065 & 0.969 & 0.941 & 1.002 & 0.464 & 0.423 & 0.509 & -0.500 & -0.515 & -0.483 \\ 
					N=196, T=10 &  &  &  &  &  &  &  &  &  &  &  &  &  &  &  &  &  &  &  &  &  &  &  &  &  &  &  &  &  &  \\ 
					True & 0.700 & -0.446 & -0.439 &  & 0.500 &  &  & -0.169 & -0.151 &  & 0.200 &  &  & 0.050 &  & 0.400 &  &  & 0.100 &  &  & 1.000 &  &  & 0.500 &  &  & -0.500 &  &  \\ 
					Estimated & 0.884 & -0.401 & -0.375 & 0.228 & 0.512 & 0.376 & 0.656 & -0.202 & -0.197 & 0.015 & 0.259 & 0.164 & 0.362 & 0.029 & 0.405 & 0.702 & 0.293 & 1.117 & 0.048 & 0.003 & 0.104 & 0.872 & 0.802 & 0.949 & 0.405 & 0.284 & 0.537 & -0.496 & -0.546 & -0.447 \\ 
					N=49, T=5 &  &  &  &  &  &  &  &  &  &  &  &  &  &  &  &  &  &  &  &  &  &  &  &  &  &  &  &  &  &  \\ 
					True & 0.700 & -0.423 & -0.380 &  & 0.500 &  &  & -0.159 & -0.135 &  & 0.200 &  &  & 0.013 &  & 0.400 &  &  & 0.100 &  &  & 1.000 &  &  & 0.500 &  &  & -0.500 &  &  \\ 
					Estimated & 0.738 & -0.349 & -0.303 & 0.363 & 0.444 & 0.364 & 0.522 & -0.211 & -0.209 & 0.013 & 0.267 & 0.198 & 0.338 & 0.000 & 0.554 & 0.692 & 0.512 & 0.869 & 0.030 & 0.003 & 0.065 & 0.963 & 0.934 & 0.989 & 0.465 & 0.420 & 0.506 & -0.500 & -0.517 & -0.485 \\ 
					N=196, T=10 &  &  &  &  &  &  &  &  &  &  &  &  &  &  &  &  &  &  &  &  &  &  &  &  &  &  &  &  &  &  \\ 
					True & 1.000 & -0.637 & -0.627 &  & 0.714 &  &  & -0.242 & -0.216 &  & 0.286 &  &  & 0.050 &  & 0.400 &  &  & 0.100 &  &  & 1.000 &  &  & 0.500 &  &  & -0.500 &  &  \\ 
					Estimated & 1.197 & -0.584 & -0.558 & 0.380 & 0.707 & 0.485 & 0.922 & -0.251 & -0.244 & 0.030 & 0.325 & 0.172 & 0.506 & 0.014 & 0.381 & 0.747 & 0.238 & 1.161 & 0.045 & 0.003 & 0.095 & 0.862 & 0.773 & 0.941 & 0.390 & 0.272 & 0.521 & -0.490 & -0.538 & -0.441 \\ 
					N=49, T=5 &  &  &  &  &  &  &  &  &  &  &  &  &  &  &  &  &  &  &  &  &  &  &  &  &  &  &  &  &  &  \\ 
					True & 1.000 & -0.604 & -0.543 &  & 0.714 &  &  & -0.228 & -0.192 &  & 0.286 &  &  & 0.013 &  & 0.400 &  &  & 0.100 &  &  & 1.000 &  &  & 0.500 &  &  & -0.500 &  &  \\ 
					Estimated & 1.121 & -0.380 & -0.330 & 0.425 & 0.488 & 0.331 & 0.626 & -0.436 & -0.419 & 0.042 & 0.552 & 0.376 & 0.682 & -0.010 & 0.501 & 0.888 & 0.635 & 1.132 & 0.046 & 0.003 & 0.107 & 0.928 & 0.885 & 0.970 & 0.468 & 0.425 & 0.516 & -0.500 & -0.518 & -0.484 \\ 
					N=196, T=10 &  &  &  &  &  &  &  &  &  &  &  &  &  &  &  &  &  &  &  &  &  &  &  &  &  &  &  &  &  &  \\ 
					True & 1.429 & -0.637 & -0.627 &  & 0.714 &  &  & -0.242 & -0.216 &  & 0.286 &  &  & 0.050 &  & 0.400 &  &  & 0.100 &  &  & 0.700 &  &  & 0.500 &  &  & -0.500 &  &  \\ 
					Estimated & 1.687 & -0.602 & -0.531 & 0.535 & 0.723 & 0.497 & 0.928 & -0.250 & -0.238 & 0.056 & 0.325 & 0.179 & 0.511 & 0.011 & 0.437 & 0.650 & 0.235 & 1.082 & 0.045 & 0.002 & 0.091 & 0.621 & 0.547 & 0.694 & 0.416 & 0.316 & 0.505 & -0.491 & -0.531 & -0.455 \\ 
					N=49, T=5 &  &  &  &  &  &  &  &  &  &  &  &  &  &  &  &  &  &  &  &  &  &  &  &  &  &  &  &  &  &  \\ 
					True & 1.429 & -0.604 & -0.543 &  & 0.714 &  &  & -0.228 & -0.192 &  & 0.286 &  &  & 0.013 &  & 0.400 &  &  & 0.100 &  &  & 0.700 &  &  & 0.500 &  &  & -0.500 &  &  \\ 
					Estimated & 1.533 & -0.490 & -0.402 & 0.612 & 0.615 & 0.497 & 0.709 & -0.325 & -0.313 & 0.064 & 0.409 & 0.319 & 0.501 & -0.022 & 0.539 & 0.692 & 0.509 & 0.902 & 0.036 & 0.003 & 0.080 & 0.668 & 0.639 & 0.693 & 0.477 & 0.445 & 0.509 & -0.500 & -0.511 & -0.488 \\ 
					N=196, T=10 &  &  &  &  &  &  &  &  &  &  &  &  &  &  &  &  &  &  &  &  &  &  &  &  &  &  &  &  &  &  \\ 
					True & 3.165 & -0.637 & -0.627 &  & 0.714 &  &  & -0.242 & -0.216 &  & 0.286 &  &  & 0.050 &  & 0.400 &  &  & 0.100 &  &  & 0.316 &  &  & 0.500 &  &  & -0.500 &  &  \\ 
					Estimated & 7.112 & -0.443 & -0.403 & 0.534 & 0.556 & 0.384 & 0.725 & -0.425 & -0.353 & 0.371 & 0.523 & 0.333 & 0.643 & 0.031 & 0.592 & 0.908 & 0.446 & 1.274 & 0.150 & 0.004 & 0.316 & 0.152 & 0.037 & 0.276 & 0.457 & 0.409 & 0.507 & -0.498 & -0.516 & -0.478 \\ 
					N=49, T=5 &  &  &  &  &  &  &  &  &  &  &  &  &  &  &  &  &  &  &  &  &  &  &  &  &  &  &  &  &  &  \\ 
					True & 3.165 & -0.604 & -0.543 &  & 0.714 &  &  & -0.228 & -0.192 &  & 0.286 &  &  & 0.013 &  & 0.400 &  &  & 0.100 &  &  & 0.316 &  &  & 0.500 &  &  & -0.500 &  &  \\ 
					Estimated & 3.960 & -0.486 & -0.405 & 0.703 & 0.611 & 0.465 & 0.741 & -0.336 & -0.302 & 0.287 & 0.422 & 0.293 & 0.561 & -0.019 & 0.570 & 0.615 & 0.400 & 0.814 & 0.116 & 0.004 & 0.197 & 0.261 & 0.162 & 0.377 & 0.490 & 0.474 & 0.506 & -0.500 & -0.506 & -0.494 \\ 
					N=196, T=10 &  &  &  &  &  &  &  &  &  &  &  &  &  &  &  &  &  &  &  &  &  &  &  &  &  &  &  &  &  &  \\ 
					True & 4.484 & -0.637 & -0.627 &  & 0.714 &  &  & -0.242 & -0.216 &  & 0.286 &  &  & 0.050 &  & 0.400 &  &  & 0.100 &  &  & 0.223 &  &  & 0.500 &  &  & -0.500 &  &  \\ 
					Estimated & 8.734 & -0.535 & -0.460 & 0.658 & 0.653 & 0.486 & 0.818 & -0.333 & -0.274 & 0.484 & 0.413 & 0.271 & 0.520 & 0.021 & 0.601 & 0.819 & 0.413 & 1.203 & 0.070 & 0.004 & 0.181 & 0.122 & 0.024 & 0.224 & 0.461 & 0.424 & 0.501 & -0.498 & -0.511 & -0.482 \\ 
					N=49, T=5 &  &  &  &  &  &  &  &  &  &  &  &  &  &  &  &  &  &  &  &  &  &  &  &  &  &  &  &  &  &  \\ 
					True & 4.484 & -0.604 & -0.543 &  & 0.714 &  &  & -0.228 & -0.192 &  & 0.286 &  &  & 0.013 &  & 0.400 &  &  & 0.100 &  &  & 0.223 &  &  & 0.500 &  &  & -0.500 &  &  \\ 
					Estimated & 4.759 & -0.574 & -0.512 & 0.776 & 0.704 & 0.602 & 0.815 & -0.248 & -0.220 & 0.362 & 0.313 & 0.203 & 0.401 & -0.018 & 0.577 & 0.491 & 0.335 & 0.640 & 0.107 & 0.004 & 0.202 & 0.214 & 0.135 & 0.296 & 0.493 & 0.481 & 0.506 & -0.500 & -0.505 & -0.495 \\ 
					N=196, T=10 &  &  &  &  &  &  &  &  &  &  &  &  &  &  &  &  &  &  &  &  &  &  &  &  &  &  &  &  &  &  \\ 
					True & 10.000 & -0.637 & -0.627 &  & 0.714 &  &  & -0.242 & -0.216 &  & 0.286 &  &  & 0.050 &  & 0.400 &  &  & 0.100 &  &  & 0.100 &  &  & 0.500 &  &  & -0.500 &  &  \\ 
					Estimated & 30.768 & -0.596 & -0.540 & 0.762 & 0.709 & 0.563 & 0.844 & -0.272 & -0.222 & 0.769 & 0.337 & 0.291 & 0.390 & 0.006 & 0.586 & 0.633 & 0.330 & 0.941 & 0.112 & 0.006 & 0.244 & 0.034 & 0.009 & 0.087 & 0.469 & 0.446 & 0.494 & -0.500 & -0.510 & -0.491 \\ 
					N=49, T=5 &  &  &  &  &  &  &  &  &  &  &  &  &  &  &  &  &  &  &  &  &  &  &  &  &  &  &  &  &  &  \\ 
					True & 10.000 & -0.604 & -0.543 &  & 0.714 &  &  & -0.228 & -0.192 &  & 0.286 &  &  & 0.013 &  & 0.400 &  &  & 0.100 &  &  & 0.100 &  &  & 0.500 &  &  & -0.500 &  &  \\ 
					Estimated & 8.800 & -0.617 & -0.571 & 0.829 & 0.746 & 0.670 & 0.828 & -0.209 & -0.174 & 0.656 & 0.263 & 0.219 & 0.298 & -0.021 & 0.591 & 0.437 & 0.317 & 0.568 & 0.042 & 0.004 & 0.103 & 0.115 & 0.083 & 0.148 & 0.496 & 0.487 & 0.503 & -0.500 & -0.504 & -0.498 \\ 
					N=196, T=10 &  &  &  &  &  &  &  &  &  &  &  &  &  &  &  &  &  &  &  &  &  &  &  &  &  &  &  &  &  &  \\
					\hline
			\end{tabular}}
			\begin{tablenotes}
				\item \tiny \textit{Note: By \textit{Population} value of $\eta^+$, $u^+$ and $v$ we mean the \textit{average} and \textit{median} values as they were generated from the Monte Carlo experiment. We report posterior means, medians and highest density intervals from posterior draws using 20,000 iterations, 10,000 burn-in and thinning equal to 5. $\hat{\mathbf{E}}(\rho)$ is the expected value of the correlation of the posterior draws of one-sided errors and spatial effects, and the \textit{Population} one-sided errors and spatial effects coming from the Monte Carlo experiment, respectively.}
			\end{tablenotes}
		\end{table} 
	\end{landscape}
	
	\begin{landscape}
		\begin{table}[p] \caption{Sampling properties of Bayes estimators: Heavy tails}\label{tab:a2}
			\centering  \resizebox{1.4\textheight}{!}{%
				\begin{tabular}{cccccccccccccccccccccccccccccc}
					\hline    &
					\multicolumn{3}{c}{$\eta^+$} &
					\multicolumn{3}{c}{$\sigma_\eta$}&
					\multicolumn{3}{c}{$u^+$}&
					\multicolumn{3}{c}{$\sigma_u$} &
					\multicolumn{2}{c}{$v$}&
					\multicolumn{3}{c}{$\sigma_v$} &
					\multicolumn{3}{c}{$\sigma_\alpha$} &
					\multicolumn{3}{c}{$\sigma_\epsilon$} &
					\multicolumn{3}{c}{$\beta_1$} &
					\multicolumn{3}{c}{$\beta_2$} \\
					\cmidrule(r){2-4}  \cmidrule(r){5-7}  \cmidrule(r){8-10}  \cmidrule(r){11-13}  \cmidrule(r){14-15}  \cmidrule(r){16-18}  \cmidrule(r){19-21}  \cmidrule(r){22-24}  \cmidrule(r){25-27}  \cmidrule(r){28-30}
					&
					Mean &
					Median &
					$\hat{\E}(\rho)$    & Mean &
					\multicolumn{2}{c}{HDI}  &   Mean &
					Median &
					$\hat{\E}(\rho)$    & Mean &
					\multicolumn{2}{c}{HDI} & 
					Median &
					$\hat{\E}(\rho)$    & Mean &
					\multicolumn{2}{c}{HDI} & Mean &
					\multicolumn{2}{c}{HDI} & Mean &
					\multicolumn{2}{c}{HDI} & Mean &
					\multicolumn{2}{c}{HDI}& Mean &
					\multicolumn{2}{c}{HDI}  \\ 
					\hline
					Population & -0.439 & -0.484 &  & 0.500 &  &  & -0.159 & -0.131 &  & 0.200 &  &  & -0.020 &  & 0.400 &  &  & 0.100 &  &  & 0.100 &  &  & 0.500 &  &  & -0.500 &  &  \\ 
					estimate & -0.380 & -0.289 & 0.664 & 0.487 & 0.387 & 0.588 & -0.209 & -0.177 & 0.533 & 0.263 & 0.198 & 0.324 & -0.057 & 0.586 & 0.526 & 0.299 & 0.745 & 0.059 & 0.003 & 0.141 & 0.103 & 0.034 & 0.149 & 0.500 & 0.474 & 0.525 & -0.500 & -0.510 & -0.489 \\ 
					N=49, T=5 &  &  &  &  &  &  &  &  &  &  &  &  &  &  &  &  &  &  &  &  &  &  &  &  &  &  &  &  &  \\ 
					Population & -0.360 & -0.334 &  & 0.500 &  &  & -0.160 & -0.129 &  & 0.200 &  &  & 0.045 &  & 0.400 &  &  & 0.100 &  &  & 0.100 &  &  & 0.500 &  &  & -0.500 &  &  \\ 
					estimate & -0.376 & -0.307 & 0.711 & 0.491 & 0.386 & 0.598 & -0.163 & -0.140 & 0.419 & 0.207 & 0.152 & 0.260 & 0.029 & 0.671 & 0.542 & 0.276 & 0.779 & 0.071 & 0.003 & 0.163 & 0.128 & 0.088 & 0.170 & 0.486 & 0.469 & 0.504 & -0.505 & -0.512 & -0.499 \\ 
					N=49, T=10 &  &  &  &  &  &  &  &  &  &  &  &  &  &  &  &  &  &  &  &  &  &  &  &  &  &  &  &  &  \\ 
					Population & -0.434 & -0.347 &  & 0.500 &  &  & -0.160 & -0.138 &  & 0.200 &  &  & -0.011 &  & 0.400 &  &  & 0.100 &  &  & 0.100 &  &  & 0.500 &  &  & -0.500 &  &  \\ 
					estimate & -0.426 & -0.387 & 0.747 & 0.525 & 0.439 & 0.607 & -0.174 & -0.155 & 0.417 & 0.219 & 0.159 & 0.278 & -0.019 & 0.264 & 0.263 & 0.139 & 0.386 & 0.060 & 0.003 & 0.143 & 0.120 & 0.061 & 0.169 & 0.494 & 0.478 & 0.512 & -0.505 & -0.512 & -0.498 \\ 
					N=100, T=5 &  &  &  &  &  &  &  &  &  &  &  &  &  &  &  &  &  &  &  &  &  &  &  &  &  &  &  &  &  \\ 
					Population & -0.344 & -0.279 &  & 0.500 &  &  & -0.165 & -0.138 &  & 0.200 &  &  & -0.024 &  & 0.400 &  &  & 0.100 &  &  & 0.100 &  &  & 0.500 &  &  & -0.500 &  &  \\ 
					estimate & -0.322 & -0.281 & 0.642 & 0.418 & 0.351 & 0.489 & -0.179 & -0.153 & 0.457 & 0.226 & 0.168 & 0.276 & -0.016 & 0.395 & 0.405 & 0.277 & 0.550 & 0.027 & 0.003 & 0.047 & 0.130 & 0.091 & 0.172 & 0.498 & 0.484 & 0.509 & -0.497 & -0.501 & -0.492 \\ 
					N=100, T=10 &  &  &  &  &  &  &  &  &  &  &  &  &  &  &  &  &  &  &  &  &  &  &  &  &  &  &  &  &  \\ 
					Population & -0.391 & -0.331 &  & 0.500 &  &  & -0.160 & -0.134 &  & 0.200 &  &  & 0.001 &  & 0.400 &  &  & 0.100 &  &  & 0.100 &  &  & 0.500 &  &  & -0.500 &  &  \\ 
					estimate & -0.391 & -0.315 & 0.686 & 0.498 & 0.431 & 0.563 & -0.162 & -0.142 & 0.425 & 0.206 & 0.154 & 0.251 & -0.028 & 0.531 & 0.501 & 0.369 & 0.624 & 0.045 & 0.004 & 0.095 & 0.140 & 0.105 & 0.179 & 0.503 & 0.495 & 0.512 & -0.502 & -0.505 & -0.499 \\ 
					N=196, T=10 &  &  &  &  &  &  &  &  &  &  &  &  &  &  &  &  &  &  &  &  &  &  &  &  &  &  &  &  &  \\ 
					\hline
			\end{tabular}}
			\begin{tablenotes}
				\item \tiny \textit{Note: By \textit{Population} value of $\eta^+$, $u^+$ and $v$ we mean the \textit{average} and \textit{median} values as they were generated from the Monte Carlo experiment. We report posterior means, medians and highest density intervals from posterior draws using 20,000 iterations, 10,000 burn-in and thinning equal to 5. $\hat{\mathbf{E}}(\rho)$ is the expected value of the correlation of the posterior draws of one-sided errors and spatial effects, and the \textit{Population} one-sided errors and spatial effects coming from the Monte Carlo experiment, respectively.}
			\end{tablenotes}
		\end{table} 
	\end{landscape}
\end{document}